# A simple but precise method for solving axisymmetric contact problems involving elastically graded materials

Markus Heß*

*Institute of Mechanics, Technical University of Berlin, Berlin, 10623, Germany*

**Abstract**

An efficient method is presented for solving axisymmetric, frictionless contact problems between a rigid punch and an elastically non-homogeneous, power-law graded half-space. Provided that the contact area is simply-connected profiles of arbitrary shape can be considered. Moreover, adhesion in the framework of the generalized JKR-theory can be taken into account. All results agree exactly with those given by three-dimensional contact theories. The method uses the fact that three-dimensional contact problems can be mapped to one-dimensional ones with a properly defined Winkler-foundation; hence, the method is to be understood as an extension of the method of dimensionality reduction (MDR). A prerequisite of its applicability forms the generality of contact stiffness regardless of the geometry of the axisymmetric profile, which is proved. All the necessary mapping rules are derived and their ease of use explained by solving contact problems based on actual examples and up to now unsolved problems.

---

**Zusammenfassung**

In diesem Paper wird eine sehr effiziente Methode zur exakten Lösung des axialsymmetrischen, reibungsfreien Kontaktproblems zwischen einem starren Indenter und einem elastisch inhomogenen Halbraum vorgestellt. Die Inhomogenität ist durch einen Elastizitätsmodul gegeben, der senkrecht zur Halbraumoberfläche nach einem Potenzgesetz zunimmt. Unter der Voraussetzung einer einfach zusammenhängenden Kontaktfläche können beliebig geformte Profile angenommen und auch Kontakte mit Adhäsion im Rahmen der JKR-Theorie gelöst werden. Die Methode nutzt die Tatsache aus, dass dreidimensionale Kontaktprobleme auf Kontaktprobleme mit einer eindimensionalen Winklerschen Bettung abgebildet werden können und ist daher als Erweiterung der Methode der Dimensionsreduktion zu verstehen. Grundvoraussetzung ihrer Anwendung bildet die Allgemeingültigkeit der Kontaktsteifigkeit unabhängig von der Geometrie des axialsymmetrischen Profils, die wir nebenbei beweisen. In diesem Beitrag werden alle notwendigen Abbildungsregeln hergeleitet und ihre einfache Anwendung zur Lösung von Kontaktproblemen anhand von aktuellen Beispielen und bisher ungelösten Problemen erläutert.

---

## 1. Introduction

*1.1. Fundamentals of functionally graded materials*

Der technologische Fortschritt einschließlich der Forderung nach immer größeren Leistungen verlangt die Entwicklung hoch technologischer Materialien. Zu dieser Klasse zählen die Functionally-Graded-Materials (FGMs), deren Anwendungsfeld von der Biomechanik, Tribologie, Optoelektronik bis hin zur Nanotechnologie reicht [1]. Im Gegensatz zu homogenen Materialien können die Werkstoffeigenschaften von FGMs optimal und zum Teil unabhängig voneinander eingestellt werden. So können hohe Korrosions- und Ermüdungsfestigkeit, Verschleißbeständigkeit, Bruchzähigkeit sowie optimierte thermische und elektrische Eigenschaften in einem Werkstoff vereint werden [2]. Im Ingenieurwesen werden solche Materialien eingesetzt, um den enormen tribologischen Beanspruchungen gerecht

---

* Corresponding author. Tel.: +49 030 314 21485

*E-mail address:* markus.hess@tu-berlin.de

zu werden und so die Sicherheit und Lebensdauer von mechanischen Bauteilen zu steigern.

Aufgrund ihrer optimierten Eigenschaften sind FGMs mit zum Teil erheblichen Herstellungskosten verbunden. Daher werden mathematische Modelle als Grundlage von numerischen Simulationen benötigt, die das Verhalten von FGMs voraussagen. In diesem Zuge erfolgt eine Homogenisierung der sehr heterogenen Mikrostruktur des Materials. FGMs werden also über Kontinua mit räumlich veränderlichen mechanischen Eigenschaften idealisiert, um die Anwendung der Kontinuumstheorie zu ermöglichen [3]. Die räumliche Veränderung der Materialeigenschaften senkrecht zur Oberfläche wird zumeist über ein Exponential- oder über ein Potenzgesetz beschrieben. Das kontaktmechanische Verhalten solcher Gradientenmaterialien musste allerdings nur bedingt neu ergründet werden. Weit bevor überhaupt von FGMs die Rede war, wurden in der Geomechanik Berechnungen angestellt, um den Einfluss eines mit der Tiefe zunehmenden Elastizitätsmoduls des Baugrunds zu studieren (siehe z. B. [4; 5]). Dabei wurden über die Jahre diverse Funktionen für den mit der Tiefe veränderlichen Elastizitätsmodul angenommen. Einen guten Überblick enthält der Artikel von Selvadurai [6] und das Buch von Aleynikov [7]. In vielen Fällen sind die Berechnungen sehr kompliziert und lassen zumeist nur numerische Lösungen zu. Die Mehrzahl der Beiträge behandelt eine exponentielle oder potenzabhängige Zunahme des Elastizitätsmoduls. Außerdem haben fast alle Beiträge gemein, dass sie einen reibungsfreien Normalkontakt voraussetzen. Bereits das Auftreten von Tangentialspannungen an der Oberfläche oder aber die Berücksichtigung von Adhäsion erschweren eine Problemlösung erheblich. Die Untersuchung von Kontakten zwischen FGMs mit Adhäsion ist u. a. aufgrund der schnellen Entwicklung in der Nanotechnologie in den Fokus gerückt. Während nicht-adhäsive Kontaktprobleme von elastisch inhomogenen Materialien in Form eines Potenzgesetzes für ausgewählte Profile lange gelöst sind [8; 9], ist der Kontakt mit Adhäsion Gegenstand aktueller Forschung [10; 11; 12]. In diesem Paper stellen wir eine alternative, sehr effiziente Methode vor, die auf exakte Lösungen der oben genannten Kontaktprobleme führt. Aufgrund ihrer Einfachheit bietet sie jedem angehenden Ingenieur den Zugang zur Lösung von adhäsiven Kontaktproblemen unter Beteiligung von Gradientenmaterialien. Dazu bedarf es lediglich Grundkenntnissen der Analysis und numerischen Simulation. Die Methode ist unter dem Namen *Methode der Dimensionsreduktion* bekannt.

*1.2. Zur Methode der Dimensionsreduktion*

Bereits für homogene, elastische oder viskoelastische Materialien kann die Lösung von Kontaktproblemen innerhalb von tribologischen Systemen mit erheblichen Schwierigkeiten verbunden sein. Einerseits muss zwischen Normal-, Tangential- und Rollkontakt unterschieden werden, andererseits müssen Adhäsion, thermische und elektrische Effekte oder aber Schmiermittel- bzw. Fremdmittelschichten berücksichtigt werden [13]. Während bei Einzelkontakten das Kontaktprofil Einfluss auf das Kontaktverhalten nimmt, muss bei einem rauen Kontakt der Mehrskalencharakter der Rauheit beachtet werden. Dies ist nur eine charakteristische Auswahl des Problemspektrums, die die Notwendigkeit von numerischen Simulationen aufdeckt, um zuverlässige Aussagen zum Reibungs- und Verschleißverhalten oder aber über den elektrischen und thermischen Widerstand treffen zu können. Aufgrund der Vielfalt und der Komplexität von Kontaktproblemen ist ihre Lösung leider nur einer kleinen forschenden Gruppe vorbehalten. Sie setzt die Kenntnis ganz unterschiedlicher dreidimensionaler Theorien voraus: die Hertzsche Kontakttheorie [14], der Tangentialkontakt nach Cattaneo [15] und Mindlin [16], das Überlagerungsprinzip von Ciavarella [17] und Jäger [18], die JKR-Theorie [19] oder aber die Theorie von Radok [20] zur Beschreibung von viskoelastischem Materialverhalten, um nur einige zu nennen. Das ist nur einer der Gründe, warum die Methode der Dimensionsreduktion (MDR) geschaffen wurde. Diese Methode bildet sämtliche Theorien zur Lösung dreidimensionaler Kontaktprobleme exakt auf eindimensionale Theorien ab. Einfacher gesprochen wird der Kontakt zwischen zwei dreidimensionalen elastischen Halbräumen auf den Kontakt zwischen einem starren, ebenen Indenter und einer eindimensionalen Winklerschen Bettung abgebildet. Die damit einhergehende Reduktion an Freiheitsgraden und damit Einsparung an Rechenzeit gegenüber der Finite-Elemente-Methode (FEM) oder der Rand-Elemente-Methode (REM) ist enorm. Anstelle von 3D- bzw. 2D-Elementen werden nur 1D-Elemente benötigt, die entlang einer Linie angeordnet sind und untereinander nicht wechselwirken.

Nachdem V. L. Popov [21] die Grundidee der MDR lieferte, haben vor allem die Arbeiten von Geike und Popov [22], Heß [23] und Pohrt et al. [24] zu ihrer Entwicklung beigetragen. Im Anschluss wurde die MDR stetig erweitert, so dass sie heute zur Lösung von Tangential- und Rollkontaktproblemen, Kontaktproblemen zwischen transversalisotropen und viskoelastischen Materialien oder aber zur Berechnung von elektrischen und thermischen Widerstän-den Anwendung findet. Auf eine Auflistung der damit verbundenen zahlreichen Beiträge möchten wir verzichten und verweisen auf das umfassende und aktuelle Werk von Popov und Heß [25]. Ergänzend sei erwähnt, dass sogar zuverlässige Verschleißberechnungen [26; 27] getätigt und Torsionskontakte mit partiellem Randgleiten [28] mittels MDR gelöst werden können. Natürlich möchten wir nicht vorenthalten, dass die Anwendbarkeit der MDR auf zufällig raue Kontakte zum Teil kontrovers diskutiert wird [29; 30]. Ihre Anwendbarkeit zur Lösung von axialsymmetrischen Kontakten mit einfach zusammenhängender Kontaktfläche steht aber völlig außer Frage und das zugehörige Anwendungsfeld ist groß. Allein auf technischem Gebiet reichen die Anwendungen von nano- bzw. mikroelektromechanischen Systemen (NEMS, MEMS) über Lager, Gelenke, Kupplungen und Getriebe bis hin zum Rad-Schiene-Kontakt. Außerdem kann die MDR zur exakten Lösung des Tangentialkontaktes zwischen rauen Oberflächen herangezogen werden, wenn die Lösung des Normalkontaktes gegeben ist [31]. In einem aktuellen Beitrag diskutiert Argatov [32] die Vor- und Nachteile der MDR. Eine Zusammenfassung aller Abbildungsregeln der MDR zur Lösung von axialsymmetrischen Kontakten findet der Leser in [33].

Alle oben genannten Beiträge beziehen sich auf den Kontakt von elastisch homogenen Halbräumen. Die Anwendbarkeit der MDR auf heterogene Medien wird einzig in einem Paper von Popov [34] diskutiert. Darin stellt Popov ausführlich alle wesentlichen Eigenschaften eines axialsymmetrischen Kontaktproblems zwischen heterogenen Medien vor, die eine Reduktion ermöglichen. Exakte und approximative Lösungsansätze mittels MDR werden gegenübergestellt und auf Beispielkontakte mit beschichteten Halbräumen angewendet. Der vorliegende Beitrag ist der Anwendung der MDR auf Kontaktprobleme mit elastisch inhomogenen Materialien gewidmet, deren Elastizitätsmodul senkrecht zur Oberfläche nach dem Potenzgesetz

$$E(z) = E_0 \left(\frac{z}{c_0}\right)^k \quad \text{with} \quad 0 \leq k < 1 \qquad (1)$$

zunimmt. Selbstverständlich darf (1) aufgrund des verschwindenden Elastizitätsmoduls an der Oberfläche nur als Idealisierung verstanden werden. Dennoch ist der Ansatz berechtigt. FEM-Rechnungen von Lee et al. [35] haben gezeigt, dass sich das prinzipielle Kontaktverhalten im Vergleich zu einem mehr realistischen, abschnittsweise definierten Gesetz kaum unterscheidet. Wie anfangs erwähnt, zeigen viele FGMs ein solches Materialverhalten. Wir werden nachweisen, dass beliebige axialsymmetrische Kontakte unter Beteiligung solcher FGMs exakt mittels MDR gelöst werden können; mit inbegriffen sind adhäsive Kontakte.

Ausgehend von der Lösung des Kontaktproblems zwischen einem flachen zylindrischen Stempel und dem elastisch inhomogenen Halbraum leiten wir in Abschnitt 2.2 die entsprechende Lösung für beliebig geformte, axialsymmetrische Kontakte her. Dabei nutzen wir einen auf Mossakovskii [36] zurückgehenden Ansatz. Es sei bemerkt, dass bereits Jin und Guo [37] auf gleiche Weise das Kontaktproblem gelöst haben. Unsere Berechnungen unterscheiden sich allerdings insofern, da wir zum einen *alle* Kontaktbeziehungen mit Hilfe von Mossakovskiis Ansatz herleiten und zum anderen diverse Formänderungen vornehmen, die den Zugang zu den Abbildungsregeln der MDR erlauben. In Abschnitt 2.3 weisen wir die Allgemeingültigkeit der Kontaktsteifigkeit für beliebig geformte axialsymmetrische Kontakte her, eine der Grundvoraussetzungen der Anwendbarkeit der MDR, die Popov [34] nannte.

Aufbauend auf den Ergebnissen aus Kapitel 2 werden in Abschnitt 3.1 sämtliche Abbildungsregeln der MDR zur Lösung des axialsymmetrischen Normalkontaktproblems zwischen einem konvex geformten starren Indenter und einem elastisch inhomogenen Halbraum generiert. In den Abschnitten 3.2 und 3.3 werden ausgewählte Kontaktprobleme mittels MDR gelöst, um ihre einfache Handhabung zu verdeutlichen. Zu den behandelten Profilen gehören die Potenzfunktion und ein konisches Profil mit parabolischer Spitze. Der Abschnitt 3.4 ist der Behandlung von Kontaktproblemen mit zylindrischen Profilen beliebigen Endprofils (konkav, konvex) gewidmet.

In Kapitel 4 erfolgt eine Erweiterung der MDR, die die Lösung von adhäsiven Kontakten im Rahmen der verallgemeinerten JKR-Theorie zulässt. In Ergänzung werden einfache Stabilitätsforderungen entwickelt, die eine direkte Berechnung des kritischen Kontaktradius und damit der maximalen Abzugskraft gestatten. Durch einen Vergleich mit aktuellen Ergebnissen [10; 12] wird die Exaktheit der Lösungen mittels MDR bestätigt. Ein Teil der Ergebnisse mittels MDR ergänzt den aktuellen Stand der Forschung.

## 2. Axisymmetric contact analysis of power-law graded materials

*2.1. Flat-punch solution*

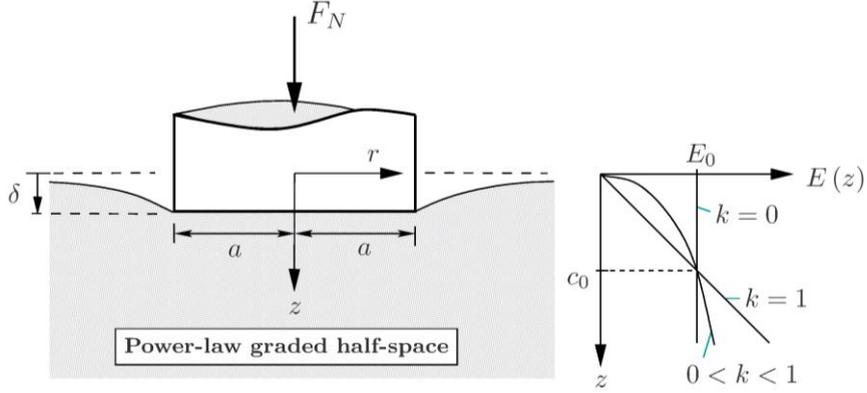

**FIG. 1: Indentation of a rigid, flat cylindrical punch into a power-law graded half-space**

Die Lösung des reibungsfreien Kontaktproblems zwischen einem flachen, zylindrischen, starren Stempel und einem elastischen Halbraum, dessen Elastizitätsmodul mit der Tiefe gemäß Gleichung (1) zunimmt (siehe FIG. 1), entnehmen wir der Arbeit von Booker et. al. [8]. Hiernach ergeben sich die Normalverschiebungen der Oberfläche des power-law graded half-space zu

$$u_z(r,a) = \begin{cases} \delta & \text{for } 0 \leq r < a \\ \delta \dfrac{2\cos\left(\dfrac{k\pi}{2}\right)}{\pi(1+k)}\left(\dfrac{a}{r}\right)^{1+k} {}_2F_1\left(\dfrac{1+k}{2}, \dfrac{1+k}{2}; \dfrac{3+k}{2}; \dfrac{a^2}{r^2}\right) & \text{for } r \geq a \end{cases}, \quad (2)$$

worin ${}_2F_1(\ldots)$ die hypergeometrische Funktion darstellt. Für die Normalspannungen an der Oberfläche des Halbraums gilt

$$p(r,a) = \begin{cases} \dfrac{h(k,\nu)E_0\delta}{\pi(1-\nu^2)c_0^k}\left(a^2-r^2\right)^{\frac{k-1}{2}} & \text{for } 0 \leq r < a \\ 0 & \text{for } r > a \end{cases}. \quad (3)$$

Aus (3) wird die Normalkraft $F_N$ in Abhängigkeit vom Kontaktradius $a$ und der Eindrücktiefe $\delta$ bestimmt

$$F_N(a,\delta) = \frac{2E_0 h(k,\nu)}{(1-\nu^2)(1+k)c_0^k}\delta a^{1+k}. \quad (4)$$

Die in den Gleichungen (3) und (4) auftretende Funktion $h(k,\nu)$ hängt vom Exponenten der elastischen Inhomogenität $k$ und der Poissonzahl $\nu$ wie folgt ab:

$$h(k,\nu) = \frac{2(1+k)\cos\left(\dfrac{k\pi}{2}\right)\Gamma\left(1+\dfrac{k}{2}\right)}{\sqrt{\pi}C(k,\nu)\beta(k,\nu)\sin\left(\dfrac{\beta(k,\nu)\pi}{2}\right)\Gamma\left(\dfrac{1+k}{2}\right)} \quad (5)$$

mit

$$C(k,\nu) = \frac{2^{1+k}\Gamma\left(\frac{3+k+\beta(k,\nu)}{2}\right)\Gamma\left(\frac{3+k-\beta(k,\nu)}{2}\right)}{\pi\,\Gamma(2+k)},$$

$$\beta(k,\nu) = \sqrt{(1+k)\left(1-\frac{k\nu}{1-\nu}\right)}.$$

Die Steifigkeit des zylindrischen Flachstempelkontaktes können wir Gleichung (4) entnehmen

$$k_N := \frac{dF_N}{d\delta} = \frac{2E_0 h(k,\nu)}{(1-\nu^2)(1+k)c_0^k}a^{1+k}. \tag{6}$$

Wir werden zu einem späteren Zeitpunkt nachweisen, dass diese Kontaktsteifigkeit für alle axialsymmetrischen Kontakte Gültigkeit besitzt.

*2.2. Axisymmetric Indentation of arbitrary profile*

Ausgehend von der Lösung des Flachstempelkontaktes, werden wir im Folgenden die Lösungen für beliebig geformte, axialsymmetrische Kontakte herleiten/entwickeln. Dabei möchten wir ausdrücklich hervorheben, dass ein Teil der Herleitungen - sogar auf gleiche Weise - bereits in der Arbeit von Jin und Guo [37] zu finden sind. Nicht nur aus Gründen des besseren Leseverständnisses erachten wir es dennoch als zwingend notwendig, diese Gleichungen hier noch einmal aufzuführen: Im Sinne des Nachweises der Möglichkeit, das axialsymmetrische Kontaktproblem auf ein 1D-Problem abzubilden, sind nämlich Umformungen einzelner Integralgleichungen erforderlich. Außerdem fehlt die Berechnung der Kontaktsteifigkeit für beliebig geformte axialsymmetrische Kontakte. Auch in dem ausgezeichneten Artikel von Jin, Guo und Zhang [11], in dem die Grundgleichungen für den axialsymmetrischen Kontakt mit Adhäsion auf elegante Art und Weise mittels des Reziprozitätssatzes von Maxwell und Betti hergeleitet werden, sind die benötigten Erweiterungen nicht enthalten.

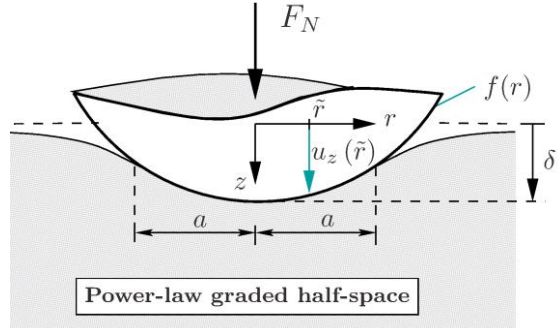

**FIG. 2: Axisymmetric contact between a rigid indenter and an elastically power-law graded half-space**

FIG. 2 zeigt ein axialsymmetrisches Kontaktproblem, welches den Randbedingungen

$$\begin{aligned} u_z(r,a) &= \delta(a)-f(r) \quad \text{für} \quad 0 \leq r \leq a \\ p(r,a) &= 0 \quad \text{für} \quad r > a \\ \tau_{zr}(r,a) &= 0 \quad \text{für} \quad r \geq 0 \end{aligned} \tag{7}$$

genügt. Für die Herleitung der Lösungen dieses Kontaktproblems nutzen wir die auf Mossakovskii [36] zurückgehende Erkenntnis aus, dass die Lösungen beliebig geformter, axialsymmetrischer Kontaktprobleme aus einer Superposition von (differenziellen) flat-punch solutions hervorgehen. Eine einfach zusammenhängende Kontaktfläche und eine konstante Poissonzahl $\nu$ werden vorausgesetzt. Wir werden die allgemeinen Lösungen in einer ganz besonderen Form entwickeln, die eine Abbildung auf ein einfaches physikalisches Ersatzsystem erlaubt. Die differenzielle

Form von (2) lautet

$$du_z(r,\tilde{a}) = \begin{cases} d\delta(\tilde{a}) & \text{for} \quad 0 \leq r < \tilde{a} \\ \dfrac{2\cos\left(\dfrac{k\pi}{2}\right)}{\pi(1+k)}\left(\dfrac{\tilde{a}}{r}\right)^{1+k} {}_2F_1\left(\dfrac{1+k}{2},\dfrac{1+k}{2};\dfrac{3+k}{2};\dfrac{\tilde{a}^2}{r^2}\right)d\delta(\tilde{a}) & \text{for} \quad r \geq \tilde{a} \end{cases} \qquad (8)$$

Die Normalverschiebung der Oberfläche innerhalb des Kontaktgebietes $(0 \leq r \leq a)$ berechnen wir mittels Integration von (8)

$$\begin{aligned} u_z(r,a) &= \int_{\delta(r)}^{\delta(a)} d\delta(\tilde{a}) + \frac{2\cos\left(\frac{1}{2}k\pi\right)}{\pi(1+k)} \int_{\delta(0)}^{\delta(r)} \left(\frac{\tilde{a}}{r}\right)^{1+k} {}_2F_1\left(\frac{1+k}{2},\frac{1+k}{2};\frac{3+k}{2};\frac{\tilde{a}^2}{r^2}\right) d\delta(\tilde{a}) \\ &= \delta(a) - \delta(r) + \frac{2\cos\left(\frac{1}{2}k\pi\right)}{\pi(1+k)} \int_0^r \left(\frac{\tilde{a}}{r}\right)^{1+k} {}_2F_1\left(\frac{1+k}{2},\frac{1+k}{2};\frac{3+k}{2};\frac{\tilde{a}^2}{r^2}\right) \frac{\partial \delta(\tilde{a})}{\partial \tilde{a}} d\tilde{a} \end{aligned} \qquad (9)$$

Nach Integration des zweiten Terms auf der rechten Seite von Gleichung (9) by parts verbleibt

$$\begin{aligned} u_z(r,a) &= \delta(a) - \delta(r) + \frac{2\cos\left(\frac{1}{2}k\pi\right)}{\pi(1+k)}\left[{}_2F_1\left(\frac{1+k}{2},\frac{1+k}{2};\frac{3+k}{2};1\right)\delta(r)\right] \\ &\quad - \frac{2\cos\left(\frac{1}{2}k\pi\right)}{\pi(1+k)} \int_0^r \left\{\frac{\partial}{\partial \tilde{a}}\left[\left(\frac{\tilde{a}}{r}\right)^{1+k} {}_2F_1\left(\frac{1+k}{2},\frac{1+k}{2};\frac{3+k}{2};\frac{\tilde{a}^2}{r^2}\right)\right]\right\} \delta(\tilde{a}) d\tilde{a} \end{aligned} \qquad (10)$$

und unter Verwendung von ${}_2F_1\left(\dfrac{1+k}{2},\dfrac{1+k}{2};\dfrac{3+k}{2};1\right) = \dfrac{\pi(1+k)}{2\cos\left(\frac{1}{2}k\pi\right)}$ sowie Ausführung der Differenziation innerhalb des Integranden

$$u_z(r,a) = \delta(a) - \frac{2\cos\left(\frac{k\pi}{2}\right)}{\pi} \int_0^r \frac{\tilde{a}^k \delta(\tilde{a})}{\left(r^2 - \tilde{a}^2\right)^{\frac{1+k}{2}}} d\tilde{a} \ . \qquad (11)$$

Ein Vergleich mit der Verschiebungsrandbedingung aus Gleichung (7) führt auf die Abelsche Integralgleichung

$$f(r) = \frac{2\cos\left(\frac{k\pi}{2}\right)}{\pi} \int_0^r \frac{\tilde{a}^k \delta(\tilde{a})}{\left(r^2 - \tilde{a}^2\right)^{\frac{1+k}{2}}} d\tilde{a} \ . \qquad (12)$$

Die Rücktransformation lautet (siehe z. B. [37])

$$\delta(\tilde{a}) = \tilde{a}^{-k} \frac{d}{d\tilde{a}} \int_0^{\tilde{a}} \frac{r f(r)}{\left(\tilde{a}^2 - r^2\right)^{\frac{1-k}{2}}} dr = \tilde{a}^{1-k} \int_0^{\tilde{a}} \frac{f'(r)}{\left(\tilde{a}^2 - r^2\right)^{\frac{1-k}{2}}} dr \ . \qquad (13)$$

Für $\tilde{a} = a$ erhalten wir aus (13) eine Bestimmungsgleichung für die Eindrücktiefe

$$\delta(a) = a^{-k} \frac{d}{da} \int_0^a \frac{r f(r)}{\left(a^2 - r^2\right)^{\frac{1-k}{2}}} dr = a^{1-k} \int_0^a \frac{f'(r)}{\left(a^2 - r^2\right)^{\frac{1-k}{2}}} dr \ . \qquad (14)$$

Die Normalverschiebung der Oberfläche außerhalb des Kontaktgebietes gewinnen wir wiederum aus (8). Unter Berücksichtigung von $r > a$ müssen wir dazu die folgende Integralgleichung lösen:

$$u_z(r,a) = \frac{2\cos\left(\frac{1}{2}k\pi\right)}{\pi(1+k)}\int_0^a \left(\frac{\tilde{a}}{r}\right)^{1+k} {}_2F_1\left(\frac{1+k}{2},\frac{1+k}{2};\frac{3+k}{2};\frac{\tilde{a}^2}{r^2}\right)\frac{\partial\delta(\tilde{a})}{\partial\tilde{a}}d\tilde{a}. \quad (15)$$

Wiederum ergibt eine Integration by parts

$$u_z(r,a) = \frac{2\cos\left(\frac{1}{2}k\pi\right)}{\pi(1+k)}\left[\underbrace{\left(\frac{a}{r}\right)^{1+k} {}_2F_1\left(\frac{1+k}{2},\frac{1+k}{2};\frac{3+k}{2};\frac{a^2}{r^2}\right)\delta(a)}_{:=C_1(a)}\right]$$

$$-\frac{2\cos\left(\frac{k\pi}{2}\right)}{\pi}\int_0^a \frac{\tilde{a}^k \delta(\tilde{a})}{\left(r^2-\tilde{a}^2\right)^{\frac{1+k}{2}}}d\tilde{a}. \quad (16)$$

Der Koeffizient $C_1(a)$ im ersten Term aus Gleichung (16) kann auch in integraler Form notiert werden, woraus

$$u_z(r,a) = \frac{2\cos\left(\frac{k\pi}{2}\right)}{\pi}\int_0^a \frac{\tilde{a}^k\left[\delta(a)-\delta(\tilde{a})\right]}{\left(r^2-\tilde{a}^2\right)^{\frac{1+k}{2}}}d\tilde{a} \quad \text{for} \quad r > a \quad (17)$$

folgt.

Die Berechnung der Normalspannungen erfolgt in gleicher Weise mit Hilfe von (3). Nur solche Flachstempel, die einen Kontaktradius $\tilde{a} > r$ besitzen, tragen zur Spannungsverteilung an der Stelle $r$ bei:

$$p(r,a) = \frac{h(k,\nu)E_0}{\pi(1-\nu^2)c_0^k}\int_{\delta(r)}^{\delta(a)}\left(\tilde{a}^2-r^2\right)^{\frac{k-1}{2}}d\delta(\tilde{a}) = \frac{h(k,\nu)E_0}{\pi(1-\nu^2)c_0^k}\int_r^a \frac{\delta'(\tilde{a})}{\left(\tilde{a}^2-r^2\right)^{\frac{1-k}{2}}}d\tilde{a}. \quad (18)$$

Völlig analog geschieht die Berechnung der Normalkraft mit Hilfe von (4):

$$F_N(a) = \frac{2E_0 h(k,\nu)}{(1-\nu^2)(1+k)c_0^k}\int_0^a \tilde{a}^{1+k}\frac{\partial\delta(\tilde{a})}{\partial\tilde{a}}d\tilde{a}. \quad (19)$$

Integration by parts and rearranging of (19) gives

$$F_N(a) = \frac{2E_0 h(k,\nu)}{(1-\nu^2)c_0^k}\int_0^a \tilde{a}^k\left[\delta(a)-\delta(\tilde{a})\right]d\tilde{a}. \quad (20)$$

Insbesondere die Gleichung (20) spielt für eine Reduktion des Kontaktproblems eine zentrale Rolle. Während wir zur Berechnung der Normalkraft von Mossakovskiis Ansatz [36] Gebrauch machen, berechnen Jin und Guo [37] die Normalkraft auf übliche Weise mittels Integration der Normalspannungen (18) über die Kontaktfläche. Dies führt zunächst auf eine von (20) abweichende Form der Bestimmungsgleichung für die Normalkraft.

*2.3. On the generality of contact stiffness*

Nachfolgend werden wir zeigen, dass alle axialsymmetrischen Kontaktprobleme zwischen einem starren Indenter und einem elastisch inhomogenen Halbraum mit dem veränderlichen Elastizitätsmodul aus Gleichung (1) die gleiche

Kontaktsteifigkeit besitzen. Dazu leiten wir die Gleichung (20) zunächst nach dem Kontaktradius $a$ ab

$$\frac{dF_N(a)}{da} = \frac{2E_0 h(k,\nu)}{(1-\nu^2)c_0^k} \left\{ \int_0^a \tilde{a}^k \frac{d\delta(a)}{da} d\tilde{a} + \left[ \tilde{a}^k \left( \delta(a) - \delta(\tilde{a}) \right) \right]_{\tilde{a}=a} \right\} = \frac{2E_0 h(k,\nu)}{(1-\nu^2)c_0^k} \frac{a^{k+1}}{k+1} \frac{d\delta(a)}{da} . \quad (21)$$

Die allgemeine Definition der Normalkontaktsteifigkeit lautet:

$$k_N := \frac{dF_N(\delta)}{d\delta} = \frac{dF_N(a)}{da} \frac{da(\delta)}{d\delta} . \quad (22)$$

Einsetzen von Gleichung (21) in Gleichung (22) führt auf die universelle Kontaktsteifigkeit

$$k_N(a) = \frac{2E_0 h(k,\nu)}{(1-\nu^2)c_0^k} \frac{a^{k+1}}{k+1} . \quad (23)$$

Die gleiche Kontaktsteifigkeit ergab sich auch für den Kontakt mit einem flachen zylindrischen Stempel, siehe Gleichung (6). Für $k=0$ folgt $h(0,\nu)=1$ und aus Gleichung (23) fällt die universelle Steifigkeit für einen axialsymmetrischen Kontakt zwischen einem starren Indenter und einem homogenen, elastischen Halbraums ab

$$k_N = 2E^* a \quad \text{with} \quad E^* = \frac{E_0}{1-\nu^2} . \quad (24)$$

Für den axialsymmetrischen Kontakt zwischen zwei elastischen Halbräumen muss der verallgemeinerte Elastizitätsmodul aus Gleichung (24) wie folgt bestimmt werden

$$\frac{1}{E^*} = \frac{1-\nu_1^2}{E_1} + \frac{1-\nu_2^2}{E_2} . \quad (25)$$

Der Nachweis für die Gültigkeit der universellen Steifigkeit für axialsymmetrische Kontakte zwischen zwei elastischen Halbräumen geht auf Pharr, Oliver und Brotzen [38] zurück. Der Nachweis erfolgte über die Integralgleichungen von Sneddon [39] und ist verglichen mit der Herleitung aus Gleichung (20) wesentlich aufwendiger. Analog zu unserem Nachweis leiteten Borodich und Keer [40] die Normalsteifigkeit für den non-slippery-contact her. Alle drei Ergebnisse sind wenig überraschend und sofort nachvollziehbar, wenn man sich an Mossakovskiis Ergebnis erinnert: Die Differenz zweier infinitesimal benachbarter Eindruckversuche mit einem Stempel beliebiger axialsymmetrischer Form ist äquivalent zu dem differenziellen Eindruck eines Flachstempels in den elastischen Halbraum. Damit muss die Kontaktsteifigkeit unabhängig von der Form des Indenters und gleich der Normalsteifigkeit beim Eindruck durch einen Flachstempel sein. Einen sehr eleganten Beweis für die Allgemeingültigkeit der Kontaktsteifigkeit liefert Popov [41]. Dieser Beweis kann zugleich als sehr anschauliche Erklärung der Erkenntnis von Mossakowskii verstanden werden.

**3. Das äquivalente eindimensionale Kontaktproblem**

*3.1. Äquivalenz des dreidimensionalen zu einem eindimensionalem Kontaktproblem*

Aus Gleichung (23) ermitteln wie das Differenzial der Kontaktsteifigkeit, welches zum Kontaktradius $\tilde{a}$ gehört

$$dk_N(\tilde{a}) = \frac{2E_0 h(k,\nu)}{(1-\nu^2)c_0^k} \tilde{a}^k d\tilde{a} = 2c_W(\tilde{a}) d\tilde{a} \quad \text{with} \quad c_W(\tilde{a}) := \frac{h(k,\nu)}{(1-\nu^2)} E(\tilde{a}) . \quad (26)$$

Der Übersichtlichkeit halber möchten wir folgende Umbenennungen vornehmen

$$\tilde{a} \rightarrow |x| \quad \text{and} \quad \delta(\tilde{a}) \rightarrow g(x) \quad (27)$$

und zugleich die neue Größe

$$u_{1D}(x) := \delta - g(x) \qquad (28)$$

einführen. Mit den neuen Bezeichnungen lauten (13) und (20) unter Verwendung von (26)

$$g(x) = |x|^{1-k} \int_0^{|x|} \frac{f'(r)}{(x^2 - r^2)^{\frac{1-k}{2}}} dr \;, \qquad (29)$$

$$F_N(a) = \int_{-a}^{a} c_W(x)[\delta - g(x)] dx = \int_{-a}^{a} c_W(x) u_{1D}(x) dx \;. \qquad (30)$$

Außerdem folgt aus (14) der Zusammenhang

$$u_{1D}(a) = 0 \;\Leftrightarrow\; \delta = g(a) \;. \qquad (31)$$

Die hervorgehobenen Gleichungen (29), (30) und (31) sind aus der exakten dreidimensionalen Theorie hervorgegangen. Diese Gleichungen sind aber gleichermaßen für ein viel einfacheres Kontaktproblem gültig, welches wir eindimensionales Äquivalent nennen. Es handelt sich dabei um den Kontakt zwischen einem starren, ebenen Profil der Form $g(x)$, welches in eine eindimensionale Schicht aus Federn der Steifigkeit $c_W(x)\Delta x$ gedrückt wird. Der Modul der elastischen Bettung (Winkler foundation) $c_W(x)$ ist dabei direkt proportional zum tiefenabhängigen Elastizitätsmodul zu wählen

$$c_W(x) = \frac{h(k,\nu)}{(1-\nu^2)} E(|x|) = \frac{h(k,\nu)}{(1-\nu^2)} E_0 \left(\frac{|x|}{c_0}\right)^k , \qquad (32)$$

wobei wir $z$ durch $|x|$ ersetzen und die Koordinate $x$ vom ersten Kontaktpunkt an zählen müssen. Zur Lösung eines axialsymmetrischen Kontaktproblems über einen äquivalenten eindimensionalen Kontakt sind demnach zwei vorbereitende Schritte notwendig:

*Step 1: Transformation of geometrical profile*

Zunächst muss die Formfunktion $\tilde{z} = f(r)$ des dreidimensionalen Indenters in ein ebenes Profil der Form $\tilde{z} = g(x)$ überführt werden (siehe FIG. 3). Das geschieht mit Hilfe der Formel (29). Da das ebene Profil zu einem eindimensionalen Kontaktproblem gehört, werden wir es im Verlauf häufig als eindimensionales Profil bezeichnen.

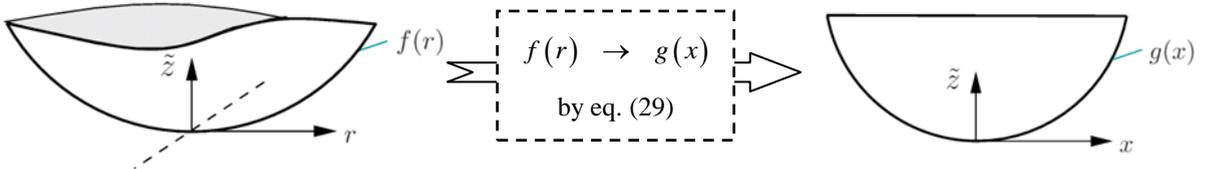

**FIG. 3: First step of MDR: Transformation of 3D-profile into 1D-profile**

Der zweite vorbereitende Schritt beinhaltet die Abbildung der elastischen Eigenschaften. Die Materialeigenschaften des elastisch inhomogenen Halbraums sind durch eine konstante Poissonzahl $\nu$ und den mit der Tiefe zunehmenden Elastizitätsmodul nach Gleichung (1) definiert. Diese Materialeigenschaften werden innerhalb der MDR durch eine

Reihe von infinitesimal benachbarten, linearen Federelementen abgebildet, deren Steifigkeit mit dem lateralen Abstand vom ersten Kontaktpunkt zunimmt (siehe rechte Seite in FIG. 5). Nach Gleichung (32) ist die Federsteifigkeit dem veränderlichen Elastizitätsmodul direkt proportional, wenn man die vertikale Koordinate $z$ durch den Betrag der lateralen Koordinate $|x|$ ersetzt. Der Proportionalitätsfaktor hängt dabei sowohl von der Poissonzahl als auch vom Exponenten der Inhomogenität ab.

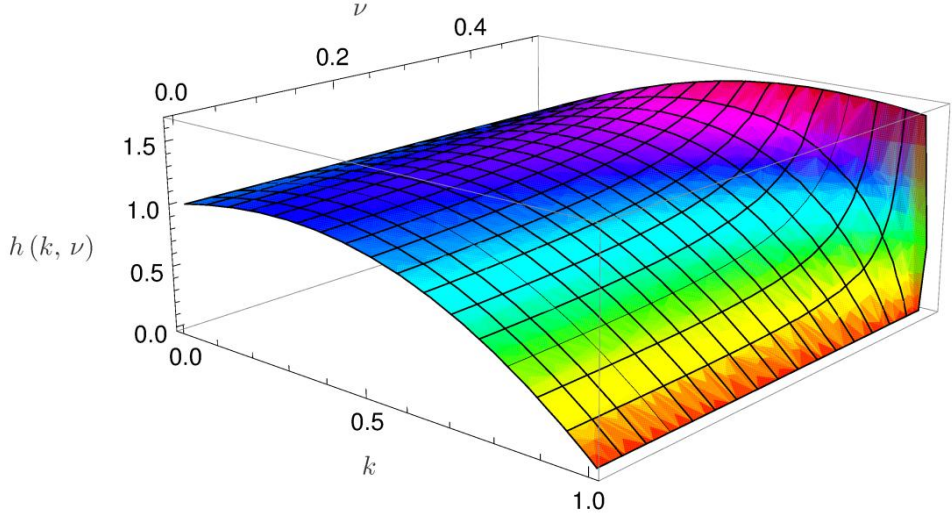

**FIG. 4: Zum Einfluss von Poissonzahl und Exponent der elastischen Inhomogenität auf die Bettungssteifigkeit**

Der Vorfaktor $h(k,\nu)$ beschreibt das charakteristische Verhalten der Steifigkeit, welches in FIG. 4 gezeigt ist. Daraus geht hervor, dass unabhängig von der Poissonzahl im homogenen Fall $k=0$ für den Vorfaktor $h(0,\nu)=1$ gilt. Die Bettungssteifigkeit entspricht dann dem verallgemeinerten Elastizitätsmodul: $c_W = E^*$. Für den linear-inhomogenen Halbraum $(k=1)$ verschwindet hingegen die Steifigkeit für alle Poissonzahlen $0 \leq \nu < 0.5$. Nur für den linear-inhomogenen, inkompressiblen Halbraums $(\nu=0.5)$ existiert eine von Null verschiedene Steifigkeit. Auf dem ersten Blick scheinen diese Ergebnisse falsch zu sein, denn abgesehen vom inkompressiblen Fall leistet die Bettung beim Eindruck mit einem beliebigen Profil keinerlei Widerstand. Beide Resultate sind aber korrekt! Denn tatsächlich ergeben sich bei einem Eindruck eines beliebigen Profils in den linear-inhomogenen, kompressiblen Halbraum unbegrenzte Verschiebungen innerhalb des Kontaktgebietes [42]. Die Ursache der Unbestimmtheit der Verschiebungen ist in dem verschwindenden Elastizitätsmoduls an der Oberfläche des Halbraums zu suchen [43]. Dass die Verschiebungen bei Inkompressibilität dagegen begrenzt sind, ist auch mit der dreidimensionalen Theorie verträglich. Dahinter verbirgt sich nämlich das so genannte Gibson-Medium [44]. Gibson wies nach, dass die Normalverschiebungen den Normalspannungen an der Oberfläche direkt proportional sind, d.h.

$$p(x,y) = \frac{3E_0}{2c_0} u_z(x,y) . \tag{33}$$

„In this case, the surface of the medium responds to changes of vertical stress like a uniform bed of springs". Gebiete an der Oberfläche, die keinen Normalspannungen unterliegen, erleiden auch keine Verschiebung! In diesem Sonderfall gilt für den Vorfaktor $h(1,0.5)=\frac{\pi}{2}$, woraus sich aus (32) für die Bettungssteifigkeit

$$c_W(x) = \frac{2\pi}{3} E_0 \frac{|x|}{c_0} \quad (34)$$

ergibt. Wir werden zu einem späteren Zeitpunkt noch zeigen, dass mittels (34) tatsächlich Normalkontakte des Gibson-Mediums exakt abgebildet werden können.

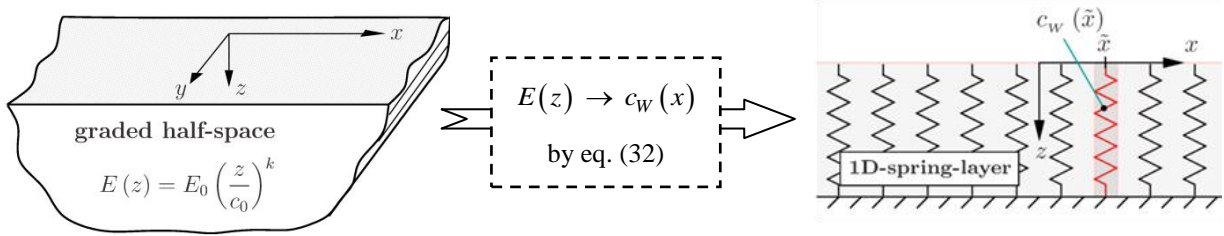

**FIG. 5: Second step of MDR: Mapping of material properties**

Nach Transformation der Profilgeometrie und Abbildung der Materialeigenschaften ist die Reduktion des dreidimensionalen Kontaktproblems aus FIG. 2 auf ein eindimensionales Kontaktproblem vollendet. Nun müssen wir lediglich das ebene Profil in die eindimensionale Winklersche Bettung drücken, was in FIG. 6 gezeigt ist.

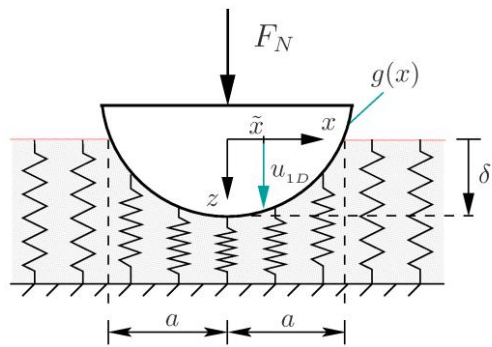

**FIG. 6: Equivalent 1D contact problem of the 3D contact problem between an axisymmetric rigid indenter of arbitrary profile and a power-law graded half-space shown in FIG. 2**

Auf diesem Wege werden wir exakt die gleichen Abhängigkeiten zwischen Normalkraft $F_N$, Eindrücktiefe $\delta$ und Kontaktradius $a$ erhalten, wie sie sich auch im dreidimensionalen Kontaktproblem ergeben würden. Aus FIG. 6 ist ersichtlich, dass die Oberflächenverschiebung der Federbettung am Kontaktrand $x = \pm a$ verschwindet, was uns unmittelbar zur Forderung aus Gleichung (31) führt. Hieraus lässt sich also der Zusammenhang zwischen Eindrücktiefe und Kontaktradius ermitteln. Die Normalkraft hingegen muss der Summe der einzelnen Federkräfte entsprechen. Sind die Abstände zwischen den Federn genügend klein gewählt $(\Delta x \to dx)$, so ergibt sich Gleichung (30) und damit die Abhängigkeit der Normalkraft vom Kontaktradius. Es sei nochmals hervorgehoben, dass sich die Steifigkeit der Federn gemäß (32) mit dem Abstand vom Zentrum des Kontaktes ändern. Beide Kontaktprobleme sind also äquivalent hinsichtlich ihres makroskopischen Verhaltens. Alle Kontaktprobleme, bei denen nur die Abhängigkeiten zwischen den makroskopischen Größen von Interesse sind, können demnach äußerst einfach mit Hilfe der MDR gelöst werden. Natürlich existieren auch zahlreiche Fragestellungen, die eine explizite Berechnung der lokalen Größen verlangen. Dazu zählen die Spannungen innerhalb der Kontaktfläche und die Oberflächennormalverschiebungen außerhalb. Es ist anschaulich klar, dass die Federspannungen im eindimensionalen Modell ungleich den Normalspannungen im dreidimensionalen Kontakt sind. Gleiches gilt für die Oberflächenverschiebungen, die im eindimensionalen Modell außerhalb des Kontaktgebietes Null sind. Dennoch gibt es eine einfache Möglichkeit, beide Größen aus dem eindimensionalen Modell exakt zu bestimmen. Dazu ist nur die Kenntnis der Oberflächenverschiebungen $u_{1D}(x)$ innerhalb der Kontaktlinie $-a \leq x \leq a$ des eindimensionalen Modells erforderlich. Unter Berück-sichtigung der Definition (28) können die Integralgleichungen (17) und (18) nämlich auch wie folgt notiert werden:

$$u_z(r,a) = \frac{2\cos\left(\frac{k\pi}{2}\right)}{\pi}\int_0^a \frac{x^k u_{1D}(x)}{(r^2-x^2)^{\frac{1+k}{2}}}dx \quad \text{for} \quad r > a \, , \tag{35}$$

$$p(r,a) = -\frac{h(k,\nu)E_0}{\pi(1-\nu^2)c_0^k}\int_r^a \frac{u'_{1D}(x)}{(x^2-r^2)^{\frac{1-k}{2}}}dx \, . \tag{36}$$

Mit den Gleichungen (35) und (36) können nunmehr auch die lokalen Größen an der Oberfläche des Kontaktproblems berechnet werden. Sollen auch die Spannungen im Halbraum bestimmt werden, so kann man die aus (36) ermittelten Oberflächenspannungen als Eingang für die Rand-Elemente-Methode heranziehen. Wie bereits erwähnt, führt uns der Sonderfall $k=0$ zur Lösung von Kontaktproblemen mit dem homogenen elastischen Halbraum. Es ist leicht zu verifizieren, dass aus den Gleichungen (29), (30), (35) und (36) dann die entsprechenden Gleichungen für den Normalkontakt der klassischen MDR hervorgehen [25].

*3.2. Example 1: indentation by a rigid indenter with a power-law profile*

Um die einfache Handhabung der MDR zu verdeutlichen, werden wir nachfolgend die Lösungen für den Kontakt zwischen einem starren, axialsymmetrischen Indenter mit einem Profil in Form einer Potenzfunktion und einem elastisch inhomogenen Halbraum herleiten.

Das axialsymmetrische Profil habe die Form

$$f(r) = A_n r^n \quad \text{with} \quad n \in \mathbb{R}^+, \, A_n = \text{const.} \, . \tag{37}$$

Zur Berechnung des 1d-Ersatzprofils müssen wir die Ableitung $f'(r) = nA_n r^{n-1}$ in Gleichung (29) einsetzen, aus der wir nach einer einfachen Rechnung

$$g(x) = \kappa(n,k)A_n|x|^n = \kappa(n,k)f(|x|) \, . \tag{38}$$

Die Gleichung (38) bedeutet anschaulich, dass das eindimensionale Profil aus einer einfachen Streckung aus dem Originalprofil hervorgeht. Der Streckfaktor ist dabei vom Exponenten der Potenzfunktion und vom Exponenten der elastischen Inhomogenität abhängig

$$\kappa(n,k) = \int_0^1 \frac{n\varsigma^{n-1}}{(1-\varsigma^2)^{\frac{1-k}{2}}}d\varsigma = \frac{n}{2}\int_0^1 t^{\frac{n}{2}-1}(1-t)^{\frac{k+1}{2}-1}dt =: \frac{n}{2}\mathrm{B}\left(\frac{n}{2},\frac{k+1}{2}\right) . \tag{39}$$

$\mathrm{B}(\cdot,\cdot)$ ist darin die vollständige Beta-Funktion. Der Streckfaktor ist in FIG. 7 graphisch veranschaulicht. Er nimmt mit steigendem Exponenten des Potenzprofils zu. Im homogenen Fall ergeben sich die bekannten Werte $\kappa(1,0) = \frac{\pi}{2}$ für den konischen und $\kappa(2,0) = 2$ für den parabolischen Indenter. Mit zunehmendem Exponenten der elastischen Inhomogenität nimmt der Streckfaktor immer mehr ab und stimmt im Grenzfall $k \to 1$ mit dem einfachen Schnitt des dreidimensionalen Profils in der $r$-$z$-Ebene überein. Letzteres gilt sogar für beliebige axialsymmetrische Profile, was unmittelbar aus (29) ersichtlich ist.

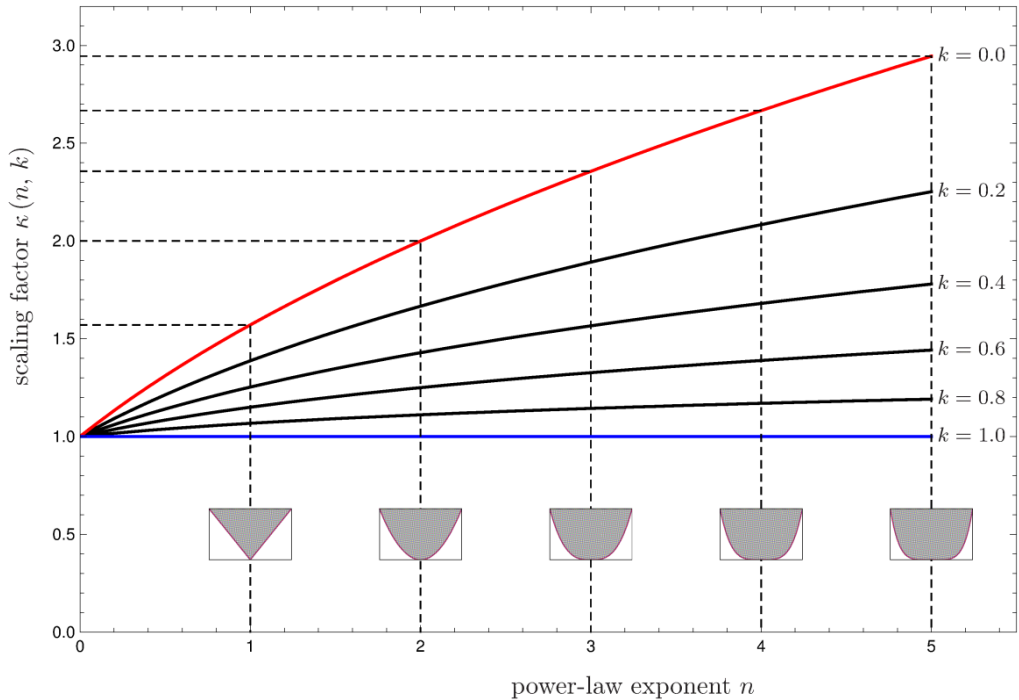

FIG. 7: Dependence of the scaling factor $\kappa$ on the power-law exponent $n$ of the profile for different exponents $k$ of the power-law graded material

Die Normalverschiebung der Oberfläche der Winklerschen Bettung stellen wir gemäß (28) auf

$$u_{1D}(x) = \delta - g(x) = \delta - \kappa(n,k) A_n |x|^n . \qquad (40)$$

Aus der Forderung (31), dass am Kontaktrand die Verschiebungen Null sind, folgt dann die Eindrücktiefe als Funktion des Kontaktradius

$$u_{1D}(a) = 0 \implies \delta = g(a) = \kappa(n,k) A_n a^n . \qquad (41)$$

Die Normalkraft erhalten wir mittels Summation aller Federkräfte. Dabei müssen wir die von der Kontaktmitte ansteigende Steifigkeit (32) beachten. Aus (30) unter Berücksichtigung von (40) und (41) ergibt sich dann

$$\begin{aligned} F_N(a) &= \int_{-a}^{a} c_W(x) u_{1D}(x) dx = 2\frac{h(k,\nu)}{(1-\nu^2)} \frac{E_0}{c_0^{\,k}} \kappa(n,k) A_n \int_0^a x^k \left(a^n - x^n\right) dx \\ &= 2\frac{h(k,\nu)}{(1-\nu^2)} \frac{E_0}{c_0^{\,k}} \kappa(n,k) A_n \frac{n}{(k+1)(n+k+1)} a^{n+k+1} \end{aligned} . \qquad (42)$$

Die in den Gleichungen (41) und (42) gegebenen makroskopischen Abhängigkeiten stimmen exakt mit den Ergebnissen der dreidimensionalen Theorie überein [9]. Für den Sonderfall $\nu = \dfrac{1}{2+k}$ wurden die Beziehungen bereits viel früher von Rostovtsev entwickelt [45].

Sind neben den Beziehungen zwischen den globalen Größen auch lokale Größen von Interesse, so können die Gleichungen (35) und (36) genutzt werden. Aus (36) erhalten wir die Normalspannungen. Differenziation von (40) nach $x$ ergibt

$$u_{1D}'(x) = -n\kappa(n,k)A_n x^{n-1}. \tag{43}$$

Einsetzen von (43) in (36) führt auf

$$\begin{aligned} p(r,a) &= \frac{h(k,\nu)E_0}{\pi(1-\nu^2)c_0^k}\int_r^a \frac{n\kappa(n,k)A_n x^{n-1}}{(x^2-r^2)^{\frac{1-k}{2}}}dx \\ &= \frac{h(k,\nu)E_0 n\kappa(n,k)A_n}{\pi(1-\nu^2)c_0^k}r^{n+k-1}\int_1^{a/r}\frac{\varsigma^{n-1}}{(\varsigma^2-1)^{\frac{1-k}{2}}}d\varsigma \\ &= \frac{h(k,\nu)E_0 n\kappa(n,k)A_n}{2\pi(1-\nu^2)c_0^k}r^{n+k-1}\left[\mathrm{B}\left(\frac{1-k-n}{2},\frac{1+k}{2}\right)-\mathrm{B}\left(\frac{r^2}{a^2},\frac{1-k-n}{2},\frac{1+k}{2}\right)\right] \end{aligned} \tag{44}$$

mit der unvollständigen Beta-Funktion

$$\mathrm{B}(z,x,y) := \int_0^z t^{x-1}(1-t)^{y-1}dt \tag{45}$$

sowie $\mathrm{B}(x,y) = \mathrm{B}(1,x,y)$. Unter Berücksichtigung von (42) können die Normalspannungen nach (44) auch auf den mittleren Duck $\bar{p}$ bezogen werden:

$$\frac{p(r,a)}{\bar{p}} = \frac{(k+1)(n+k+1)}{4}\left(\frac{r}{a}\right)^{n+k-1}\left[\mathrm{B}\left(\frac{1-k-n}{2},\frac{1+k}{2}\right)-\mathrm{B}\left(\frac{r^2}{a^2},\frac{1-k-n}{2},\frac{1+k}{2}\right)\right]. \tag{46}$$

Die Normalverschiebungen der Oberfläche außerhalb des Kontaktgebietes berechnen wir mittels (35). Unter Berücksichtigung von (40) und (41) ergibt sich

$$\begin{aligned} u_z(r,a) &= \frac{2\cos\left(\frac{k\pi}{2}\right)\kappa(n,k)A_n}{\pi}r^n\int_0^{a/r}\frac{\varsigma^k\left[\left(\frac{a}{r}\right)^n-\varsigma^n\right]}{(1-\varsigma^2)^{\frac{1+k}{2}}}d\varsigma \\ &= \frac{\cos\left(\frac{k\pi}{2}\right)\kappa(n,k)A_n}{\pi}r^n\left[\left(\frac{a}{r}\right)^n\mathrm{B}\left(\frac{a^2}{r^2},\frac{1+k}{2},\frac{1-k}{2}\right)-\mathrm{B}\left(\frac{a^2}{r^2},\frac{1+k+n}{2},\frac{1-k}{2}\right)\right] \end{aligned} \tag{47}$$

Unter Beachtung der Identität

$$\mathrm{B}(z,x,y) = \frac{z^x}{x}{}_2F_1(x,1-y;1+x;z) \tag{48}$$

sind das genau die in [11] gegebenen Verschiebungen. Um die Abhängigkeit der Verschiebungen von dem Exponenten der Inhomogenität einmal aufzuzeigen, greifen wir den parabolischen Kontakt heraus. Die auf die Eindrücktiefe des Hertzschen Kontaktes normierten Verschiebungen lauten dann

$$\frac{u_z(r,a)}{2A_2 a^2} = \begin{cases} \dfrac{\kappa(2,k)}{2}-\dfrac{1}{2}\left(\dfrac{r}{a}\right)^2 & \text{for } r \leq a \\ \dfrac{\cos\left(\frac{k\pi}{2}\right)\kappa(2,k)}{2\pi}\left[\mathrm{B}\left(\dfrac{a^2}{r^2},\dfrac{1+k}{2},\dfrac{1-k}{2}\right)-\left(\dfrac{r}{a}\right)^2\mathrm{B}\left(\dfrac{a^2}{r^2},\dfrac{3+k}{2},\dfrac{1-k}{2}\right)\right] & \text{for } r > a \end{cases} \tag{49}$$

Die Verschiebungen nach (49) sind in FIG. 8 graphisch veranschaulicht. Die Kurve für $k=0$ stimmt exakt mit der Lösung für den Hertzschen Kontakt (siehe z. B. [46]) überein:

$$\frac{u_z(r,a)}{2A_2a^2} = \frac{1}{\pi}\left[\left(2-\left(\frac{r}{a}\right)^2\right)\arcsin\left(\frac{a}{r}\right)+\sqrt{\left(\frac{r}{a}\right)-1}\right] \quad \text{for} \quad r>a. \tag{50}$$

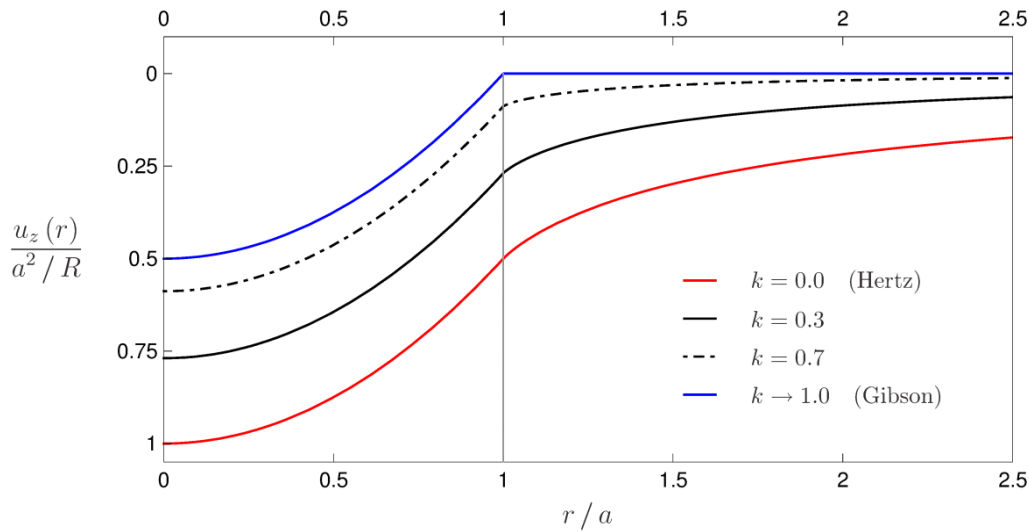

**FIG. 8: Comparison of the surface displacements bei einem parabolic contact für verschiedene Inhomogenitäten**

Mit ansteigendem Exponenten der Inhomogenität verringern sich die Verschiebungen außerhalb des Kontaktgebietes. Im Grenzfall $k \rightarrow 1$ verhält sich das Medium wie eine Winklersche Bettung. Hierbei ist allerdings anzumerken, dass es in Einklang mit den Ergebnissen von Gibson nur für $\nu = 1/2$ eine endliche Verschiebung innerhalb des Kontaktgebietes gibt. Im eindimensionalen Modell finden wir diesen Effekt in der bereits angesprochenen Tatsache wieder, dass es nur bei Inkompressibilität eine von Null verschiedene Federsteifigkeit gibt. In allen anderen Fällen bei $k \rightarrow 1$ verschwindet die Federsteifigkeit, so dass einer angelegten Normalkraft nie das Gleichgewicht gehalten werden kann (siehe (42)).

*3.3. Example 2: indentation by a rigid conical indenter with a parabolic tip*

Lassen Sie uns nun zur Untersuchung eines Kontaktes kommen, der nach bestem Wissen noch nicht in der Literatur behandelt wurde: Der Kontakt eines konischen Indenters mit abgerundeter Spitze mit einem power-law graded half-space (see FIG. 9). Ciavarella [47] untersuchte den Normal-und Tangentialkontakt zwischen diesem Stempel und dem homogenen elastischen Halbraum; ein Teil seiner Lösungen müsste daher auch aus unserer Berechnung hervorgehen.

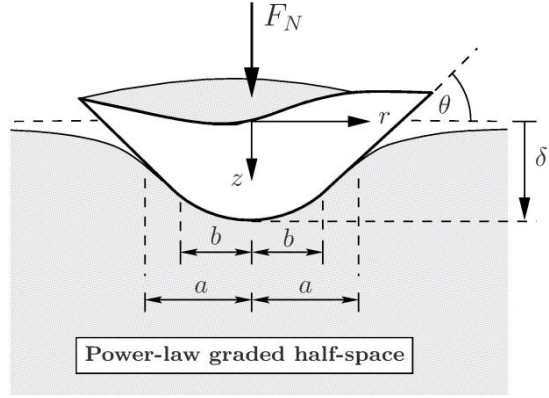

**FIG. 9: Indentation of a conical punch with a rounded tip into a power-law graded half-space**

Der Einfachheit halber seien nur die Zusammenhänge zwischen Kontaktradius, Eindrücktiefe und Normalkraft von Interesse. Das dreidimensionale Profil ist durch die Formfunktion

$$f(r) = \begin{cases} \dfrac{\tan\theta}{2b}r^2 & \text{for} \quad 0 \leq r \leq b \\ \dfrac{\tan\theta}{2}(2r-b) & \text{for} \quad r > b \end{cases} \tag{51}$$

gegeben. Der Radius $b$ ist dabei ein charakteristisches Abmaß für den parabolischen Abschnitt. Zur Berechnung des Ersatzprofils benötigen wir die Ableitung des Profils nach der radialen Koordinate

$$f'(r) = \begin{cases} \dfrac{\tan\theta}{b}r & \text{for} \quad 0 \leq r \leq b \\ \tan\theta & \text{for} \quad r > b \end{cases}. \tag{52}$$

Eingesetzt in (29) ergibt sich

$$g(x) = \begin{cases} |x|^{1-k} \displaystyle\int_0^{|x|} \dfrac{\tan\theta}{(x^2-r^2)^{\frac{1-k}{2}}} \dfrac{r}{b} dr & \text{for} \quad |x| \leq b \\ |x|^{1-k} \left[ \displaystyle\int_0^{b} \dfrac{\tan\theta}{(x^2-r^2)^{\frac{1-k}{2}}} \dfrac{r}{b} dr + \int_b^{|x|} \dfrac{\tan\theta}{(x^2-r^2)^{\frac{1-k}{2}}} dr \right] & \text{for} \quad |x| > b \end{cases}. \tag{53}$$

Die Berechnung für den Bereich $|x| \leq b$ können wir uns ersparen und das Ergebnis Gleichung (38) entnehmen. Lediglich der Bereich $|x| > b$ erfordert eine neue Rechnung, was nach einigen Umformungen auf

$$g(x) = \begin{cases} \dfrac{\tan\theta}{(1+k)b} x^2 & \text{for} \quad |x| \leq b \\ \dfrac{\tan\theta}{(1+k)b} x^2 \left[ 1 - \left(1 - \left(\dfrac{b}{x}\right)^2\right)^{\frac{k+1}{2}} \right] + \dfrac{\tan\theta}{2}|x| \left[ \mathrm{B}\left(\dfrac{1}{2}, \dfrac{1+k}{2}\right) - \mathrm{B}\left(\dfrac{b^2}{x^2}, \dfrac{1}{2}, \dfrac{1+k}{2}\right) \right] & \text{for} \quad |x| > b \end{cases} \tag{54}$$

führt. In normierter Form und im Vergleich zum Originalprofil sind die zu (54) zugehörigen Graphen für ausgewählte Exponenten der Inhomogenität in FIG. 10 geplottet. Offensichtlich ist die Abweichung des 1D-Profils vom 3D-Profil im Rahmen der klassischen MDR – also bei Indentierung des homogenen Halbraums – am größten. Mit zunehmenden Exponenten der Inhomogenität nähert sich das Profil immer mehr dem ebenen Schnittbild des 3D-Profils an, mit dem es für $k \to 1$ sogar übereinstimmt.

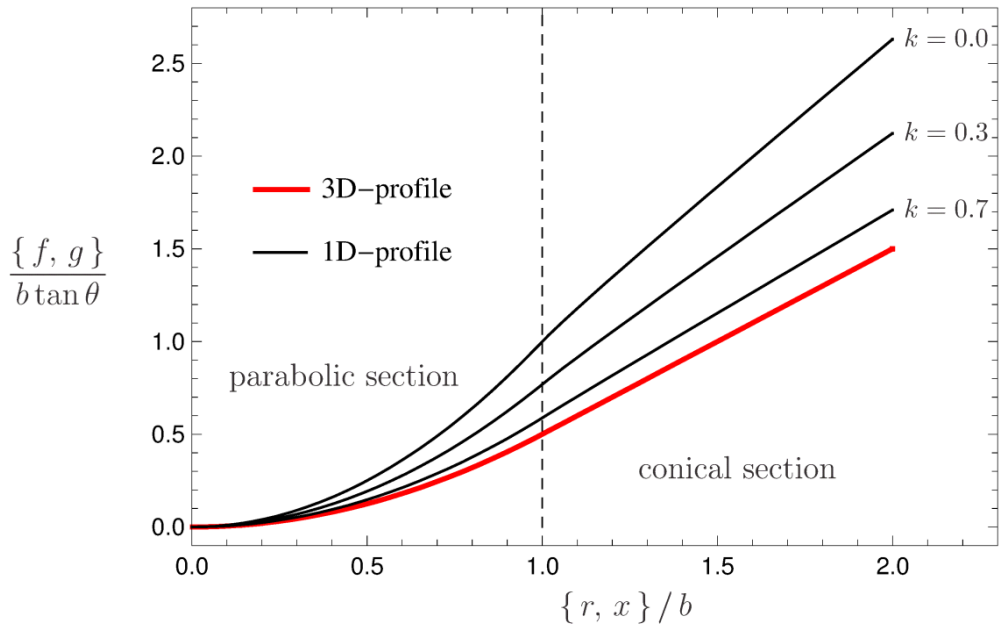

**FIG. 10: Dreidimensionales und eindimensionales Profile für verschiedene k im Vergleich; Für k=1 stimmen 1D-Profil und 3D-Profil überein**

Nun ist es ein leichtes, die Eindrücktiefe zu finden. Dazu müssen wir nach (31) lediglich $x = a$ in (54) einsetzen:

$$\delta(a) = \begin{cases} \dfrac{\tan\theta}{(1+k)b} a^2 & \text{for} \quad a \leq b \\ \dfrac{\tan\theta}{(1+k)b} a^2 \left[ 1 - \left(1 - \left(\dfrac{b}{a}\right)^2\right)^{\frac{k+1}{2}} \right] + \dfrac{\tan\theta}{2} a \left[ B\left(\dfrac{1}{2}, \dfrac{1+k}{2}\right) - B\left(\dfrac{b^2}{a^2}, \dfrac{1}{2}, \dfrac{1+k}{2}\right) \right] & \text{for} \quad a > b \end{cases}. \quad (55)$$

Wenn wir in FIG. 10 die Koordinate $x$ durch den Kontaktradius $a$ ersetzen, dann geben die Ersatzprofile genau die Änderung der Eindrücktiefe mit dem Kontaktradius wieder. Die Eindrücktiefe nimmt wie erwartet mit steigendem Exponenten der Inhomogenität ab. Das Ergebnis ist also abhängig von $k$, unabhängig hingegen von der charakteristischen Länge $c_0$.

Drücken wir das eindimensionale Profil aus (54) bis zu einer Eindrücktiefe gemäß (55) in die eindimensionale Winklersche Bettung ein, wird sich exakt die Normalkraft des 3D-Kontaktproblems einstellen. Dabei müssen wir die vom Kontaktmittelpunkt nach außen hin zunehmende Bettungssteifigkeit gemäß (32) beachten und nach (30) einfach alle Federkräfte aufsummieren. Solange wir nur so weit eindrücken, dass der Kontaktbereich nicht über den parabolischen Abschnitt hinausgeht $(a \leq b)$, wird sich das Ergebnis aus (42) für $n = 2$ einstellen. Bei weiterem Eindruck $(a > b)$ kommt der Einfluss des äußeren, konischen Abschnitts zum Tragen. Einsetzen der Verschiebung der Winklerschen Bettung in die Bestimmungsgleichung (30) für die Normalkraft führt auf

$$F_N(a) = 2\int_0^a c_W(x) u_{1D}(x) dx$$

$$= 2\frac{h(k,\nu)}{(1-\nu^2)}\frac{E_0}{c_0^{\,k}}\left\{\int_0^a x^k \delta dx - \int_0^b x^k g_I(x) dx - \int_b^a x^k g_{II}(x) dx\right\} \quad \text{for} \quad a > b$$

(56)

Die Indizes $I, II$ bezeichnen darin die beiden, definierten Abschnitte der Funktion $g(x)$ aus Gleichung (54). Aufgrund der Unhandlichkeit wird die explizite Integration von Gleichung (56) per Hand auf den Anhang verschoben. Dabei sei angemerkt, dass sich bei bekanntem eindimensionalen Profil die numerische Umsetzung und Lösung von Gleichung (30) immer (!) extrem einfach gestaltet; Gleichung (56) stellt also keine Hürde dar! Die Ergebnisse für die Normalkraft sind in FIG. 11 für ein inkompressibles Material und einer charakteristischen Länge $c_0 = b$ visualisiert. Dabei wurde zwischen kleinen und großen Normalkräften unterschieden. Um bei nicht allzu großen Kräften einen bestimmten Kontaktradius $a \approx 0,...,2b$ einzustellen, muss beim Eindruck in den homogenen Halbraum eine größere Normalkraft aufgebracht werden als bei einem graded material (see FIG. 11 left). Im Bereich $2b < a < 4b$ wechselt dieses Verhalten. Nun muss das graded material für den gleichen Kontaktradius mit einer größeren Kraft beansprucht werden als der homogene Halbraum (see FIG. 11 right). Dieses Verhalten ist völlig plausibel und unabhängig von der Form des Indenters, da mit zunehmender Eindrücktiefe der mit der Tiefe ansteigende Elastizitätsmodul immer mehr zur Geltung kommt. Die Position des Übergangsbereiches ist dabei streng abhängig von der Wahl der charakteristischen Tiefe $c_0$. Wählen wir beispielsweise $c_0 = 0.1b$, so reagiert das graded medium bereits bei sehr kleinen Kräften „steifer" als das homogene Medium. Ist die charakteristische Tiefe $c_0$, in welcher der Elastizitätsmodul $E_0$ vorherrscht, dagegen sehr groß, so setzt sich der Effekt erst bei sehr großen Normalkräften durch. Das geht aus FIG. 12 hervor, in welcher die Fälle $c_0 = 0.1b$ und $c_0 = 10b$ gegenübergestellt sind. Man könnte den Einfluss der charakteristischen Tiefe und des Exponenten der Inhomogenität an dieser Stelle noch genauer untersuchen, in dem man mittels Gleichung (36) die Normalspannungen berechnet. Das ist aber nicht Sinn dieser Arbeit, in der die Vermittlung der einfachen Handhabung der MDR zur Lösung von Kontaktproblemen mit graded materials im Vordergrund steht.

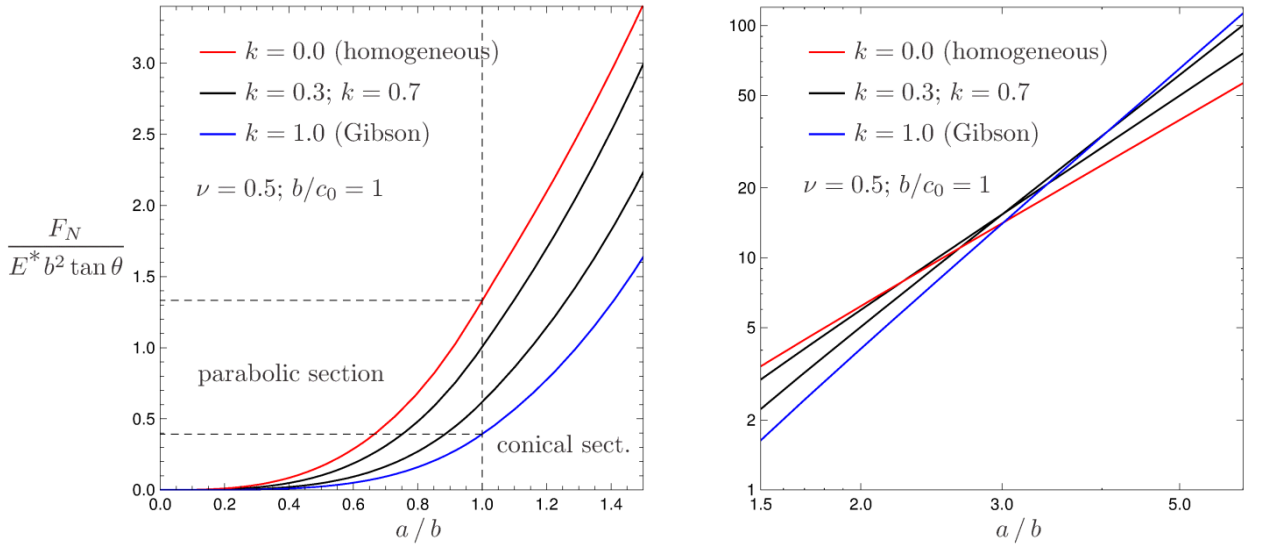

**FIG. 11: Nahfeld (left) und Fernfeld (right) of the non-dimensional load for a conical indenter with a parabolic tip as a function of $b/a$ ; es wurde $b/c_0 = 1$ angenommen und zwischen ausgewählten Exponenten der Inhomogenität unterschieden**

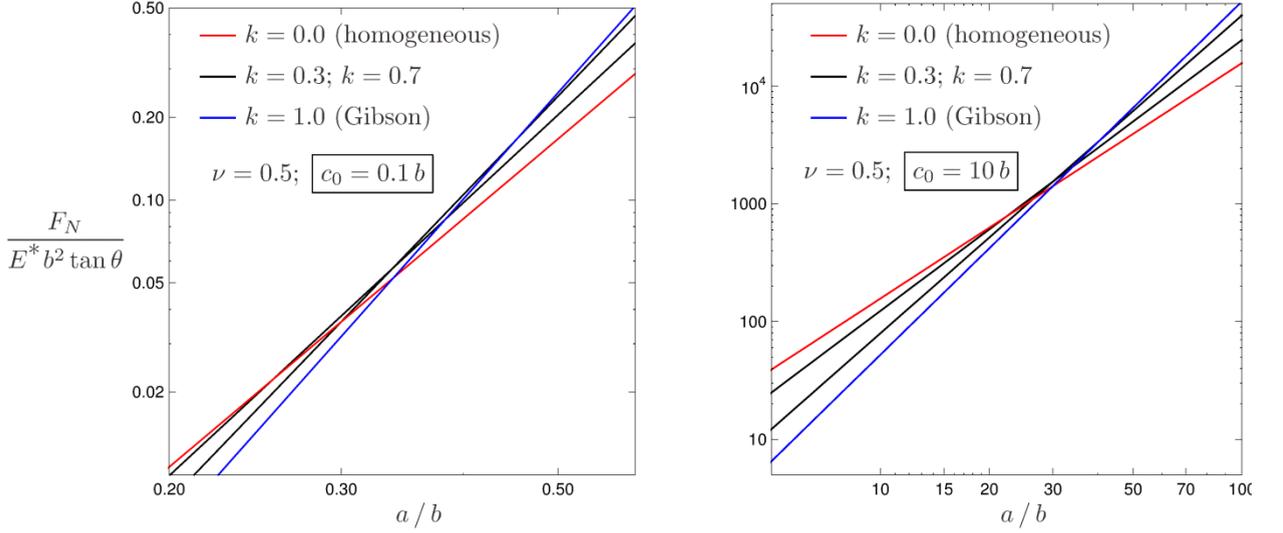

**FIG. 12: Behaviour of the non-dimensional load for a conical indenter with a parabolic tip for two different characteristic depths**

*3.4. The treatment of non-smooth contact profiles*

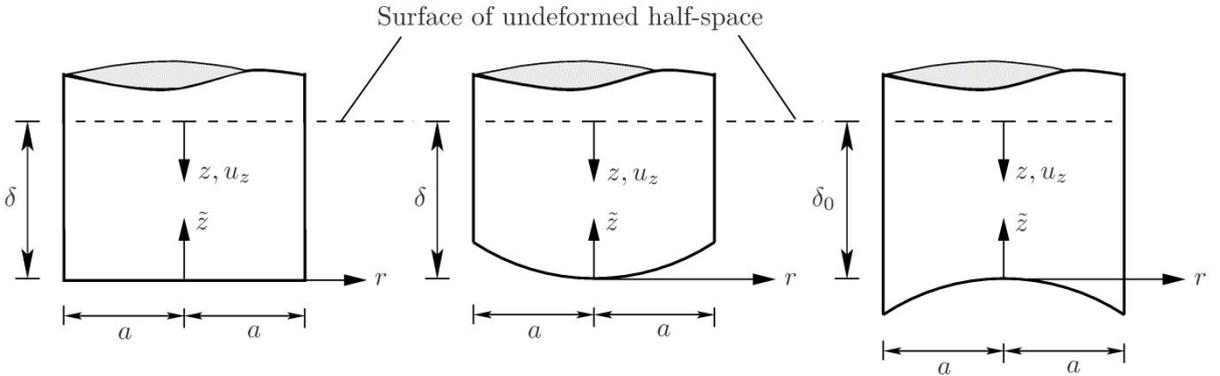

**FIG. 13: Flat cylindrical punch (left); cylindrical punch with a convex tip (middle) and with a concave tip (right)**

Für die gezeigten, zylindrischen Kontaktprofile aus FIG. 13 sind die Eindrücktiefe und der Kontaktradius unabhängig voneinander, vorausgesetzt ein vollständiger Kontakt liegt vor. Die Gleichung (31) der MDR-Regeln ist daher hinfällig. Die Definition der Verschiebungen im 1D-Modell nach Gleichung (28) ist allerdings nach wie vor gültig; gleiches gilt sowohl für die Umrechnungsformel des Ersatzprofils (29) als auch für die Berechnungsformel der Normalkraft (30), was wir nachfolgend erläutern. Gleichung (28) kann wie folgt umgeschrieben werden

$$u_{1D}(x) := \delta - g(x) = \underbrace{u_{1D}(a)}_{\substack{\text{rigid body}\\\text{translation}}} + \underbrace{g(a) - g(x)}_{\substack{\text{smooth contact-}\\\text{profile}}}. \tag{57}$$

Aus (57) ist ersichtlich, dass sich die Verschiebung gegenüber der eines glatten Profils lediglich um die Starrkörpertranslation $u_{1D}(a)$ unterscheidet. Dieser Starrkörperfreiheitsgrad beschreibt aber gerade den Eindruck eines flachen zylindrischen Stempels (see FIG. 13 left). Dass die MDR auch einem solchen Kontakt genügt, ist trivial. Aus $f(r) = 0$ folgt $g(x) = 0$ und somit

$$u_{1D}(x) = \begin{cases} \delta & \text{for} \quad |x| \leq a \\ 0 & \text{for} \quad |x| > a \end{cases}. \tag{58}$$

Zur Abbildung des Flachstempelkontaktes ist daher keinerlei Geometrietransformation erforderlich. Einsetzen von (58) in (30) liefert

$$F_N(a,\delta) = 2\delta \int_0^a c_W(x)dx \tag{59}$$

und führt unter Berücksichtigung der Bettungssteifigkeit (32) auf das exakte Ergebnis aus Gleichung (4). In gleicher Weise können mittels Anwendung von Gleichung (35) die Verschiebungen der Oberfläche außerhalb des Kontaktgebietes nach Gleichung (2) verifiziert werden. Die Berechnung der Druckverteilung im Kontaktgebiet nach Gleichung (36) ist hingegen nur dann anwendbar, wenn wir die Distributionstheorie zur Hilfe nehmen und zugleich die obere Integrationsgrenze über den Kontaktradius hinaus erweitern, also

$$p(r,a) = -\frac{h(k,\nu)E_0}{\pi(1-\nu^2)c_0^k} \int_r^\infty \frac{u'_{1D}(x)}{(x^2-r^2)^{\frac{1-k}{2}}}dx. \tag{60}$$

Die Verschiebungen der Winklerschen Bettung nach (58) können alternativ in der Form

$$u_{1D}(x) = \delta\big[\mathrm{H}(x+a)-\mathrm{H}(x-a)\big] \quad \text{with} \quad x \in \mathbb{R} \tag{61}$$

notiert werden. Darin bezeichnet $\mathrm{H}(\ldots)$ die Heaviside-Funktion. Die Ableitung von (61) nach $x$ ist

$$u'_{1D}(x) = \delta\big[\delta_D(x+a)-\delta_D(x-a)\big] \quad \text{with} \quad x \in \mathbb{R}, \tag{62}$$

worin $\delta_D(\ldots)$ die Delta-Distribution angibt. Einsetzen von (62) in Gleichung (60) unter Berücksichtigung der Filtereigenschaft der Delta-Distribution ergibt

$$\begin{aligned} p(r,a) &= -\frac{h(k,\nu)E_0}{\pi(1-\nu^2)c_0^k} \int_r^\infty \frac{\delta\big[\delta_D(x+a)-\delta_D(x-a)\big]}{(x^2-r^2)^{\frac{1-k}{2}}}dx \\ &= \frac{h(k,\nu)E_0\delta}{\pi(1-\nu^2)c_0^k (a^2-r^2)^{\frac{1-k}{2}}} \end{aligned} \tag{63}$$

Das Resultat aus (63) stimmt wiederum mit dem exakten Ergebnis aus Gleichung (3) überein. Es sei angemerkt, dass wir auch ohne Nutzung der Distributionstheorie auf die Druckverteilung schließen können. Dazu muss (36) lediglich um einen Starrkörpertranslationsanteil ergänzt werden

$$p(r,a) = -\frac{h(k,\nu)E_0}{\pi(1-\nu^2)c_0^k}\left[\int_r^a \frac{u'_{1D}(x)}{(x^2-r^2)^{\frac{1-k}{2}}}dx - \frac{u_{1D}(a)}{(a^2-r^2)^{\frac{1-k}{2}}}\right]. \tag{64}$$

Zum Abschluss sei kurz auf die zylindrischen Kontaktprofile mit konvexer und konkaver Spitze aus FIG. 13 eingegangen. Als Sonderfall einer konvexen Spitze sei eine parabolische Spitze mit dem Krümmungsradius $R$ angenommen. Dann lautet die Verschiebungsrandbedingung

$$u_z(r) \;=\; \delta - f(r) \;=\; \delta - \frac{r^2}{R} \quad \text{for} \quad r \leq a \; . \tag{65}$$

Nach Profiltransformation folgt daraus die Verschiebung der Winklerschen Bettung im Kontaktgebiet

$$u_{1D}(x) \;=\; \delta - g(x) \;=\; \delta - \frac{2}{k+1}\frac{x^2}{R} \quad \text{for} \quad |x| \leq a \; . \tag{66}$$

Aus der Schreibweise nach Gleichung (57)

$$u_{1D}(x) \;=\; u_{1D}(a) + g(a) - g(x) \;=\; \underbrace{\delta - \frac{2}{k+1}\frac{a^2}{R}}_{=u_{\text{Flat}}} + \underbrace{\frac{2}{k+1}\frac{a^2 - x^2}{R}}_{=u_{\text{Hertz}}} \quad \text{for} \quad |x| \leq a \tag{67}$$

können die zu superponierenden Anteile $u_{\text{Flat}}$ und $u_{\text{Hertz}}$ abgelesen werden. Für beide Anteile kennen wir bereits die Lösung, so dass wir die Gesamtlösung nur zusammensetzen müssen.

Bei einer konkaven Spitze - wir nehmen eine parabolische Spitze an - müssen wir beachten, dass die maximale Verschiebung und damit die Eindrücktiefe am Kontaktrand vorliegt. Die Regeln der MDR besitzen aber nur dann Gültigkeit, wenn die Verschiebung im Mittelpunkt des Kontaktgebietes im 3D- und 1D-Kontakt übereinstimmen. Daher ist anstelle der Eindrücktiefe $\delta$ die Mittenverschiebung $\delta_0$ zu nutzen (see FIG. 13 right). Die Verschiebungsrandbedingung lautet

$$u_z(r) \;=\; \delta_0 - f(r) \;=\; \delta_0 + \frac{r^2}{R} \quad \text{for} \quad r \leq a \tag{68}$$

und analog dem konvexen Profil folgt für die eindimensionalen Verschiebungen

$$u_{1D}(x) \;=\; \underbrace{\delta_0 + \frac{2}{k+1}\frac{a^2}{R}}_{=u_{\text{Flat}}} - \underbrace{\frac{2}{k+1}\frac{a^2 - x^2}{R}}_{=u_{\text{Hertz}}} \quad \text{for} \quad |x| \leq a \; . \tag{69}$$

Auch die Lösung dieses Kontaktproblems könnten wir aufgrund der Gültigkeit des Superpositionsprinzips sofort hinschreiben, worauf an dieser Stelle verzichtet wird. Dieser Abschnitt zeigt nicht nur auf, wie non-smooth Kontaktprofile im Rahmen der MDR behandelt werden können, sondern liefert gleichzeitig die Grundlage für die Abbildung von Kontakten mit Adhäsion, die Gegenstand des folgenden Kapitels sind.

**4. Die exakte Abbildung von adhäsiven Kontakten**

*4.1. Herleitung der Abbildungsregeln*

Nachfolgend werden wir erläutern, wie die verallgemeinerte Adhäsionstheorie von Johnson, Kendall und Roberts [19] angewandt auf Kontakte mit dem power-law graded half-space exakt abgebildet wird. Dabei wollen wir konvex geformte Profile voraussetzen. Im Grunde genommen bedarf es dazu keines weiteren Nachweises, denn im Rahmen der JKR-Theorie stellt der adhäsive Kontakt eine einfache Superposition des nicht-adhäsiven Kontaktes mit einer Starrkörpertranslation dar. In den Kap. 3.1 und 3.4 haben wir aufgezeigt, dass sowohl der Eindruck beliebiger, konvex geformter Profile als auch der Eindruck eines zylindrischen Flachstempels exakt abgebildet werden können. Bei gleichen Wirkungsgebieten gilt dies natürlich auch für deren Überlagerung. Die Superposition im Rahmen der MDR ist in FIG. 14 veranschaulicht. Linkerhand ist der nicht-adhäsive Andruckvorgang zu sehen, charakterisiert über die Normalkraft $F_{n.a.}$ und Eindrücktiefe $\delta_{n.a.}$. Bei einer anschließenden Entlastung gehen wir zunächst davon aus, dass alle kontaktierenden Federn am Ersatzprofil haften. Bei konstantem Kontaktradius $a$ erfahren also alle Federn die gleiche zusätzliche Verschiebung, die wir direkt an den Randfedern ablesen können. Das Gleichgewicht des Kontaktes mit Adhäsion ist allerdings erst dann gefunden, wenn die zusätzliche Verschiebung und damit die Längenände-rung der Randfedern einen ganz bestimmten Wert annimmt, den wir $\Delta\ell_{\max}$ nennen (see FIG. 14, rechts).

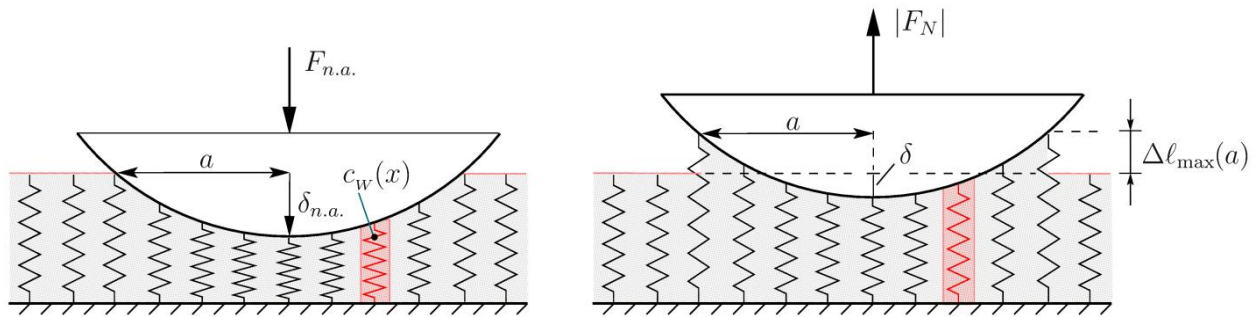

**FIG. 14: Qualitative Darstellung des Andruck- und Abziehvorgangs innerhalb der MDR zur Abbildung der verallgemeinerten JKR-Theorie für Kontakte mit einem power-law graded half-space**

Johnson, Kendall und Roberts nutzten die Forderung aus, dass im Gleichgewicht des adhäsiven Kontaktes die Gesamtenergie minimal wird. Zugleich wiesen sie auf die Gleichwertigkeit ihres Ansatzes mit jenem von Griffith hin [48; 49]. Letzteren wollen wir dazu gebrauchen, um die Gleichgewichtslänge $\Delta\ell_{\max}$ zu ermitteln. Nach dem energetischen Bruchkriterium von Griffith muss die bei der Verminderung der Kontaktfläche freigesetzte mechanische Energie der zur Bildung der neuen Oberfläche notwendigen Energie entsprechen:

$$\mathcal{G} = \Delta\gamma \,. \tag{70}$$

Darin meint $\mathcal{G}$ die elastische Energiefreisetzungsrate, die über die elastische Energie $U_E$ wie folgt berechnet werden kann

$$\mathcal{G} = \left.\frac{\partial U_E}{\partial A}\right|_\delta = \left.\frac{1}{2\pi a}\frac{\partial U_E}{\partial a}\right|_\delta \,. \tag{71}$$

$\Delta\gamma$ hingegen bezeichnet die Dupré'sche Adhäsionsenergie, die gemäß

$$\Delta\gamma := \gamma_1 + \gamma_2 - \gamma_{12} \tag{72}$$

von den Oberflächenenergien $\gamma_1$ und $\gamma_2$ der beiden Körper sowie der Grenzflächenenergie $\gamma_{12}$ abhängt. Zur Berechnung von $\mathcal{G}$ benötigen wir die elastische Energie. Da das eindimensionale System die Normalkraft als Funktion der Eindrücktiefe $F_N(\delta)$ exakt wiedergibt, muss auch die von der Winklerschen Bettung gespeicherte elastische Energie gleich der Formänderungsenergie des power-law graded half-space sein. Damit gilt

$$U_E = \frac{1}{2}\int_{-a}^{a} c_W(x) u_{1D}^2(x)\,dx = \int_0^a c_W(x) u_{1D}^2(x)\,dx \,, \tag{73}$$

mit dem Modul der elastischen Bettung nach (32) und der eindimensionalen Verschiebung nach Gleichung (28) mit einer noch unbekannten Eindrücktiefe $\delta$. Mit Hilfe von Gleichung (73) kann nun die Energiefreisetzungsrate nach Gleichung (71) ermittelt

$$\mathcal{G} = \frac{1}{2\pi a} c_W(a) u_{1D}^2(a) \tag{74}$$

und in das Griffith-Kriterium (70) eingesetzt werden. Als Ergebnis verbleibt

$$u_{1D}(a) \stackrel{!}{=} -\Delta \ell_{\max}(a) \quad \text{mit} \quad \Delta \ell_{\max}(a) := \sqrt{\frac{2\pi \Delta \gamma a}{c_W(a)}} \ . \tag{75}$$

Im Gleichgewicht des adhäsiven Kontaktes muss die Verschiebung der Randfedern an den Stellen $x = \pm a$ also genau den in Gleichung (75) gegebenen Wert annehmen. Mit Hilfe der Forderung (75) kann die Eindrücktiefe $\delta$ als Funktion des Kontaktradius $a$ für den Kontakt mit Adhäsion auf einfache Weise aus dem Ersatzmodell bestimmt werden. Gleichung (75) nimmt den Platz der Forderung (31) für den Kontakt ohne Adhäsion ein. Alle weiteren Abbildungsregeln der MDR bleiben gültig. So wird die Normalkraft aus Gleichung (30) bestimmt, während die Verschiebungen außerhalb des Kontaktgebietes aus Gleichung (35) und die Druckverteilung im Kontaktgebiet (aufgrund der überlagerten Translation) aus Gleichung (60) oder alternativ aus Gleichung (64) hervorgehen.

Einsetzen des Moduls der Winklerschen Bettung (32) an der Stelle $x = a$ in die Forderung (75) ergibt für die Längenänderung der Randfedern

$$\Delta \ell_{\max}(a) = \sqrt{\frac{2\pi \Delta \gamma a^{1-k} c_0^{k}}{E^* h(k,\nu)}} \ . \tag{76}$$

Gleichung (76) kann als Ablösekriterium für die Randfedern verstanden werden. Abgesehen vom Gibson-Medium $(k \to 1, \nu = 1/2)$ ist das Ablösekriterium offensichtlich vom aktuellen Kontaktradius abhängig. Schauen wir uns dazu nochmal FIG. 14 an und gehen davon aus, dass die Abbildung des Profils $f(r) \to g(x)$ nach (29) bereits ausgeführt wurde. Dann wird linkerhand das eindimensionale Profil unter der Last $F_{n.a.}$ in die Winklersche Bettung gedrückt und es stellt sich die Eindrücktiefe $\delta_{n.a.}$ ein. Die Randfedern befinden sich dann gerade noch im ungespannten Zustand. Nehmen wir weiter an, dass alle in Kontakt befindlichen Federn am Profil haften und wir die Normalkraft sukzessive reduzieren. Dann werden vom Kontaktrand nach innen laufend immer mehr Federn auf Zug beansprucht. Der Kontaktradius bleibt zunächst unverändert. Erreicht die Längenänderung der Randfedern den maximal zulässigen Wert $\Delta \ell_{\max}(a)$ (see FIG. 14 rechterhand), so liegt ein indifferenter Zustand zwischen Haften und Abreißen vor, der exakt mit dem Gleichgewichtszustand des dreidimensionalen, adhäsiven Kontaktes übereinstimmt. Bei weiterer Steigerung der Zugkraft werden die äußeren Federn abspringen. Im Rahmen einer numerischen Umsetzung könnte man die elastische Bettung durch eine äquidistante Anordnung von Federn modellieren. Beim Abspringen der Randfedern verringert sich der Kontaktradius dann um $\Delta x$. Bei konstanter Normalkraft wird sich die Eindrücktiefe ein wenig verringern und die Längenänderung in den neuen Randfedern $\Delta \ell(a - \Delta x)$ entsprechend vergrößern. Ist diese wiederum kleiner als die neue zulässige Längenänderung $\Delta \ell_{\max}(a - \Delta x)$ nach (76), so kann die Zugkraft weiter erhöht werden, bis sich eben jener Wert einstellt. Ein weiterer Gleichgewichtszustand des Kontaktes mit Adhäsion ist gefunden.

Im Sonderfall $k = 0$ folgt aus (76)

$$\Delta \ell_{\max}(a) = \sqrt{\frac{2\pi \Delta \gamma a}{E^*}} \ . \tag{77}$$

Dies ist das Ablösekriterium zur exakten Abbildung von adhäsiven Normalkontakten zwischen elastisch homogenen Halbräumen und geht auf Heß [23; 50] zurück.

*4.2. Einfache Stabilitätskriterien zur Berechnung der kritischen Größen*

Dass wir den Kontakt mit Adhäsion exakt mittel MDR abbilden können, haben wir im letzten Abschnitt gezeigt. Aus der Kenntnis der Normalkraft und der Eindrücktiefe als Funktion des Kontaktradius können wir mittels Extremwertuntersuchungen natürlich auch die kritischen Größen berechnen. Dennoch stellt sich die Frage, ob wir die kritischen Größen innerhalb des eindimensionalen Systems nicht auch durch einfachere Kriterien filtern können. Da der Rand des adhäsiven Kontaktes als Riss im Modus I aufgefasst werden kann, muss für stabiles Gleichgewicht die Bedingung

$$\frac{\partial \mathcal{G}}{\partial A} = \frac{1}{2\pi a}\frac{\partial \mathcal{G}}{\partial a} > 0 \tag{78}$$

erfüllt sein. Die Auswertung von Gleichung (78) muss, je nachdem welche Größe gesteuert wird, entweder bei konstanter Normalkraft (fixed load) oder konstanter Eindrücktiefe (fixed grips) erfolgen. Unter Beachtung von (74) ergibt eine kurze Rechnung

$$\frac{u_{1D}(a)}{a} > \frac{2}{1-k}\frac{\partial u_{1D}(a)}{\partial a} . \tag{79}$$

Die Ableitung auf der rechten von (79) bei konstanter Normalkraft können wir am einfachsten aus Gleichung (30) ermitteln. Wenn wir diese Gleichung bei konstanter Normalkraft nach $a$ differenzieren, verbleibt

$$\left.\frac{\partial u_{1D}(a)}{\partial a}\right|_{F_N} = -(k+1)\frac{u_{1D}(a)}{a} - \frac{\partial g(a)}{\partial a} . \tag{80}$$

Die Ableitung bei konstanter Eindrücktiefe ist hingegen trivial:

$$\left.\frac{\partial u_{1D}(a)}{\partial a}\right|_{\delta} = -\frac{\partial g(a)}{\partial a} . \tag{81}$$

Einsetzen von (80) bzw. (81) in (79) unter Berücksichtigung der Gleichgewichtsrelation (75) liefert nach wenigen Umformungen

$$\frac{\Delta \ell_{\max}(a)}{a} < \tilde{C}(k)\frac{\partial g(a)}{\partial a} \quad \text{with} \quad \tilde{C}(k) := \begin{cases} \dfrac{2}{3+k} & \text{for} \quad F_N = const. \\ \dfrac{2}{1-k} & \text{for} \quad \delta = const. \end{cases} \tag{82}$$

Die kritischen Größen gehören zum grenzstabilen Zustand. Dazu muss in (82) das "<" durch ein "=" ersetzt werden:

$$\boxed{\frac{\Delta \ell_{\max}(a_c)}{a_c} = \tilde{C}(k)\left.\frac{\partial g(a)}{\partial a}\right|_{a_c} \quad \text{with} \quad \tilde{C}(k) := \begin{cases} \dfrac{2}{3+k} & \text{for} \quad F_N = const. \\ \dfrac{2}{1-k} & \text{for} \quad \delta = const. \end{cases}} \tag{83}$$

Mit Hilfe von Gleichung (83) ist eine einfache Bedingung gefunden, aus welcher der kritische Kontaktradius $a_c$ hervorgeht. Danach hat die Steigung des Profils am Kontaktrand entscheidenden Einfluss auf den kritischen Kontaktradius.

*4.3. Example 1: adhesive contact between a power-law graded half-space and a rigid indenter with a power-law profile*

Den nicht-adhäsiven Kontakt zwischen einem elastisch inhomogenen Halbraum und einem Indenter, dessen Profil eine Potenzfunktion darstellt, haben wir bereits in Abschnitt 3.2 untersucht. Da sich die Berechnungen für den adhäsiven Kontakt einzig durch die neue Forderung (75) unterschieden, können wir die meisten Ergebnisse übernehmen. Die Transformation des Profils ergab (see eq. (37), (38))

$$f(r) = A_n r^n \quad \Rightarrow \quad g(x) = \kappa(n,k) A_n |x|^n \quad \text{with} \quad \kappa(n,k) = \frac{n}{2}\mathrm{B}\left(\frac{n}{2},\frac{k+1}{2}\right) . \tag{84}$$

Die Verschiebung der Winklerschen Bettung ist über Gleichung (28) definiert und durch (40) gegeben

$$u_{1D}(x) = \delta(a) - \kappa(n,k) A_n |x|^n \ . \tag{40}$$

Aus der Forderung (75) folgt dann die Eindrücktiefe als Funktion des Kontaktradius

$$u_{1D}(a) \stackrel{!}{=} -\Delta\ell_{\max}(a) \quad \Rightarrow \quad \delta(a) = \kappa(n,k) A_n a^n - \Delta\ell_{\max}(a) \ . \tag{85}$$

Nach Gleichung (85) ist von der Lösung im nicht-adhäsiven Fall lediglich $\Delta\ell_{\max}(a)$ abzuziehen. Einsetzen von (76) führt letztendlich auf

$$\delta(a) = \kappa(n,k) A_n a^n - \sqrt{\frac{2\pi \Delta\gamma a^{1-k} c_0^k}{E^* h(k,\nu)}} \ . \tag{86}$$

Zur Berechnung der Normalkraft wenden wir Gleichung (30) an, wobei wir die Eindrücktiefe mit Hilfe von Gleichung (85) bzw. (86) eliminieren

$$F_N(a) = \int_{-a}^{a} c_W(x) \underbrace{\left[\delta(a) - g(x)\right]}_{:=u_{1D}(x)} dx = \underbrace{2\int_0^a c_W(x) \kappa(n,k) A_n (a^n - x^n) dx}_{:=F_{n.a.}} - 2\int_0^a c_W(x) \Delta\ell_{\max}(a) dx \ . \tag{87}$$

Nach Gleichung (87) ist von der Lösung des Kontaktes ohne Adhäsion $F_{n.a.}(a)$ gemäß (42) einfach ein Ausdruck abzuziehen, der völlig unabhängig von der Geometrie immer gleich ist. Aus (87) folgt

$$F_N(a) = 2\frac{h(k,\nu) E^*}{c_0^k} \kappa(n,k) A_n \frac{n}{(k+1)(n+k+1)} a^{n+k+1} - \sqrt{\frac{8\pi \Delta\gamma h(k,\nu) E^*}{(k+1)^2 c_0^k}} a^{\frac{k+3}{2}} \ . \tag{88}$$

Zur Berechnung der kritischen Kontaktradien (unter fixed-load bzw. fixed-grips) nutzen wir das Stabilitätskriterium (83). Dazu benötigen wir nur die Steigung des Ersatzprofils am Kontaktrand

$$\frac{\partial g(a)}{\partial a} = n\kappa(n,k) A_n a^{n-1} \ . \tag{89}$$

Das Stabilitätskriterium lautet damit

$$\frac{\Delta\ell_{\max}(a_c)}{a_c} = \tilde{C}(k) n\kappa(n,k) A_n a_c^{n-1} \tag{90}$$

und unter Beachtung der Gleichgewichtslängenänderung (76) folgen nach einfachen Umformungen die kritischen Kontaktradien

$$a_c = \left(\frac{2\pi \Delta\gamma c_0^k}{h(k,\nu) E^* \tilde{C}(k)^2 \kappa(n,k)^2 n^2 A_n^2}\right)^{\frac{1}{2n+k-1}} \ . \tag{91}$$

Die kritischen Eindrücktiefen und die kritischen Normalkräfte erhalten wir nach Einsetzen von (91) in (86) und (88) und einigen langweiligen Umformungen, die wir hier nicht explizit aufführen

$$\delta_c := \delta(a_c) = \left[\left(\frac{2\pi \Delta\gamma c_0^k}{h(k,\nu) E^*}\right)^n \left(\frac{1}{\tilde{C}(k) n\kappa(n,k) A_n}\right)^{1-k}\right]^{\frac{1}{2n+k-1}} \left(\frac{1-n\tilde{C}(k)}{n\tilde{C}(k)}\right) \ , \tag{92}$$

$$F_c := F_N(a_c) = 2\left(\frac{1-\tilde{C}(k)(n+k+1)}{\tilde{C}(k)(n+k+1)(k+1)}\right)\left[(2\pi\Delta\gamma)^{n+k+1}\left(\frac{c_0^k}{h(k,\nu)E^*}\right)^{2-n}\left(\frac{1}{\tilde{C}(k)n\kappa(n,k)A_n}\right)^{k+3}\right]^{\frac{1}{2n+k-1}}. \quad (93)$$

Die Eindrücktiefe und die Normalkraft als Funktion vom Kontaktradius gemäß (86) und (88) stimmen exakt mit den in [11] gegeben Größen überein. Gleiches gilt für die kritischen Größen (91) - (93) für den Fall „fixed-load", d. h. unter Berücksichtigung von $\tilde{C}(k) = \frac{2}{3+k}$. Mit dieser Vorgabe liefert der Betrag von (93) die maximale Abzugskraft und (92) die zugehörige Eindrücktiefe. Bei Überschreiten der maximalen Abzugskraft löst sich der Kontakt vollständig. Liegt hingegen ein weggesteuerter Eindruckversuch vor, so kann die Eindrücktiefe weiter verkleinert werden, bis $\delta_c$ unter Beachtung von $\tilde{C}(k) = \frac{2}{1-k}$ erreicht ist.

Den adhäsiven Kontakt mit einem homogenen elastischen Halbraum bei kraftgesteuertem Eindruck untersuchten Yao, Gao [51]. Es ist leicht nachweisbar, dass diese Lösungen mit obigen Gleichungen für $k = 0$ übereinstimmen.

Beziehen wir den Kontaktradius, die Eindrücktiefe und die Normalkraft auf ihre kritischen Größen, so erhalten wir folgende dimensionslose Gleichungen

$$\tilde{\delta}(\tilde{a}) = \frac{1}{1-\tilde{C}(k)n}\left(\tilde{a}^n - n\tilde{C}(k)\tilde{a}^{\frac{1-k}{2}}\right) \quad \text{with} \quad \tilde{\delta} := \frac{\delta}{\delta_c} \quad \text{and} \quad \tilde{a} := \frac{a}{a_c}, \quad (94)$$

$$\tilde{F}_N(\tilde{a}) = \frac{1}{1-\tilde{C}(k)(n+k+1)}\left[\tilde{a}^{n+k+1} - \tilde{C}(k)(n+k+1)\tilde{a}^{\frac{k+3}{2}}\right] \quad \text{with} \quad \tilde{F}_N := \frac{F_N}{F_c}. \quad (95)$$

Nach (94) und (95) sind in normierter Form die Eindrücktiefe und die Normalkraft nur vom Exponenten der elastischen Inhomogenität $k$ und vom Exponenten des Potenzprofils $n$ abhängig. Weder die charakteristische Tiefe $c_0$ noch irgendwelche Materialkennwerte beeinflussen die dimensionslose Form. Für den Sonderfall eines parabolischen Kontaktes $n = 2$ wurde diese interessante Erkenntnis erst kürzlich gemacht [12].

Der Vollständigkeit halber möchten wir die kritischen Größen zu fixed-load und fixed-grips noch ins Verhältnis setzen:

$$\begin{aligned}
a_{c,\delta} &= \left(\frac{1-k}{3+k}\right)^{\frac{2}{2n+k-1}} a_{c,F}, \\
\delta_{c,\delta} &= \frac{1-k-2n}{3+k-2n}\left(\frac{1-k}{3+k}\right)^{\frac{1-k}{2n+k-1}} \delta_{c,F}, \\
F_{c,\delta} &= -\frac{1+3k+2n}{1-k-2n}\left(\frac{1-k}{3+k}\right)^{\frac{3+k}{2n+k-1}} F_{c,F}.
\end{aligned} \quad (96)$$

Für den homogenen Halbraum $k = 0$ stimmen die Relationen in Gleichung (96) mit den von Heß [23] gegebenen überein.

*4.4. Example 2: adhesive contact between a power-law graded half-space and a rigid indenter with a parabolic profile*

Selbstverständlich stellt die Betrachtung des parabolischen Kontaktes nur einen Sonderfall des vorher behandelten Potenzprofils dar. Deshalb verzichten wir auf die Wiederholung der Anwendung der MDR-Regeln und möchten stattdessen aufbauend auf obigen Ergebnissen eine kleine Analyse des Einflusses der geometrischen und der physika-lischen Kennwerte auf die Eindrücktiefe und Normalkraft anschließen. Für $n=2$, $\kappa(2,k)=2/(k+1)$ und $A_2=1/2R$ ergeben sich aus (86) und (88)

$$\frac{\delta(a)}{R} = \frac{1}{k+1}\left(\frac{a}{R}\right)^2 - \sqrt{\frac{2\pi(1-\nu^2)}{h(k,\nu)}}\left(\frac{RE_0}{\Delta\gamma}\right)^{-1/2}\left(\frac{R}{c_0}\right)^{-n/2}\left(\frac{a}{R}\right)^{\frac{1-k}{2}}, \qquad (97)$$

$$\frac{F_N}{\frac{3}{2}\pi R\Delta\gamma} = \frac{8h(k,\nu)}{3\pi(k+1)^2(k+3)(1-\nu^2)}\frac{E_0 R}{\Delta\gamma}\left(\frac{R}{c_0}\right)^k\left(\frac{a}{R}\right)^{k+3} - \frac{4}{3(k+1)}\sqrt{\frac{2h(k,\nu)}{\pi(1-\nu^2)}}\left(\frac{E_0 R}{\Delta\gamma}\right)^{1/2}\left(\frac{R}{c_0}\right)^{k/2}\left(\frac{a}{R}\right)^{\frac{k+3}{2}}. \quad (98)$$

Außerdem geht aus Gleichung (93) unmittelbar hervor, dass für $n=2$ die kritischen Normalkräfte unabhängig von den elastischen Parametern $E_0$, $\nu$ und unabhängig von der charakteristischen Tiefe $c_0$ sind

$$F_c = \left(\frac{1-\tilde{C}(k)(3+k)}{\tilde{C}(k)^2(3+k)}\right)2\pi\Delta\gamma R = \begin{cases} -\dfrac{k+3}{2}\pi\Delta\gamma R & \text{for} \quad F_N = const. \\ -\dfrac{(1-k)(5+3k)}{2(3+k)}\pi\Delta\gamma R & \text{for} \quad \delta = const. \end{cases} \qquad (99)$$

Nach (99) vergrößert sich die maximale Abzugskraft im Vergleich zum homogenen Fall mit steigendem $k$; im gleichen Zuge verringert sich der Betrag der Abzugskraft unter fixed-grips Bedingungen. Für ausgewählte Parameter ist diese Abhängigkeit in FIG. 15 festgehalten.

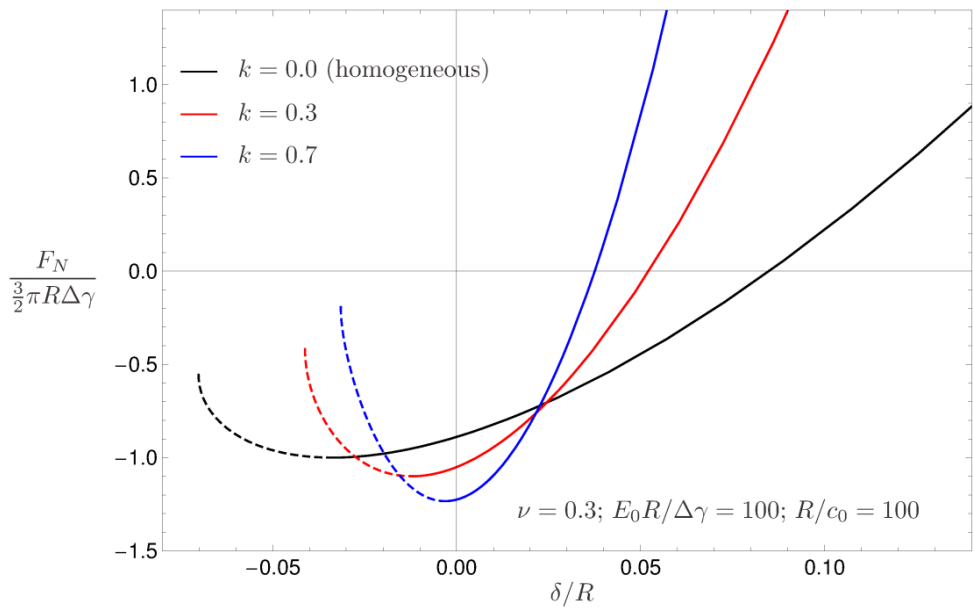

**FIG. 15: Normalkraft als Funktion der Eindrücktiefe in normierter Darstellung für den Eindruck eines parabolischen Indenters in den power-law graded half-space für verschiedene Exponenten der Inhomogenität**

Dass die kritischen Kräfte $F_c$ unter fixed-load und fixed-grips bei konstantem $k$ unabhängig von der charakteristischen Tiefe sind, zeigt dagegen FIG. 16. Auch eine Veränderung der dimensionslosen Größe $E_0 R/\Delta\gamma$ würde nichts daran ändern. In FIG. 16 ist zudem ersichtlich, dass mit zunehmender charakteristischer Tiefe $c_0$ die $F_N$-$\delta$-Kurven breiter auseinander laufen. Dieser Effekt ist plausibel, wenn man bedenkt, dass der Modul der elastischen Bettung bei Zunahme von $c_0$ kleiner wird.

Abgesehen von einer gesonderten Untersuchung der fixed-grips-Bedingungen wurde der parabolische, adhäsive Kontakt mit einem power-law graded half-space bereits von Chen, et. al. [10] untersucht. Dort sind zahlreiche Parameterstudien zu finden, die unter anderem den Verlauf der Spannungen und eine gesonderte Untersuchung des Gibson Mediums mit einschließen. Es sei angemerkt, dass die wenigen exemplarischen Ergebnisse, die wir für den parabolischen Kontakt herausgegriffen haben, exakt mit jenen übereinstimmen, die Chen, et. al. angeben. Im Unterschied zu Chen et. al. wurden unsere Ergebnisse allerdings mittels der Methode der Dimensionsreduktion gewonnen.

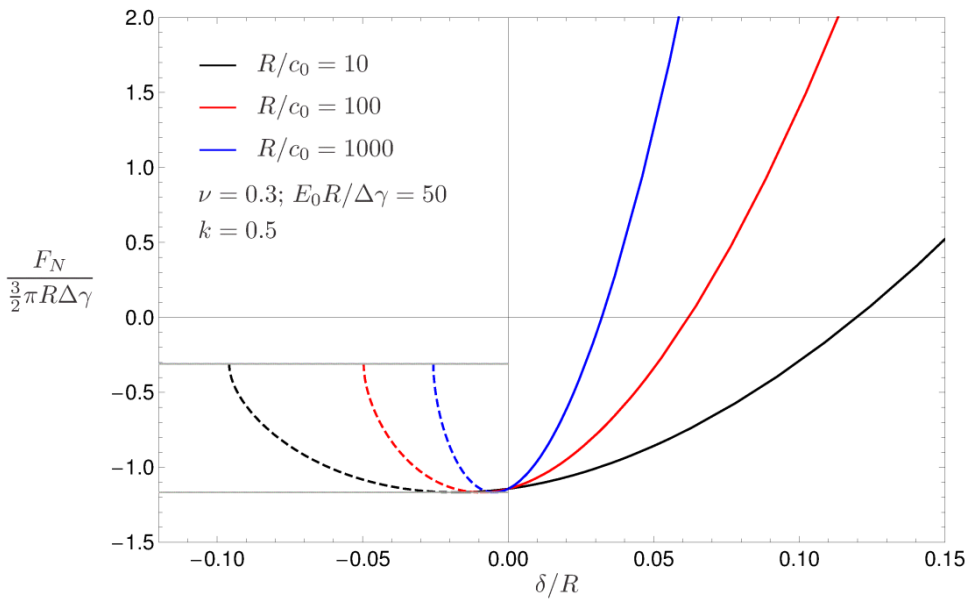

**FIG. 16: Normalkraft als Funktion der Eindrücktiefe in normierter Darstellung für den Eindruck eines parabolischen Indenters in den power-law graded half-space mit $k = 0.5$ für verschiedene charakteristische Längen $c_0$.**

## 5. Concluding remarks and outlook

In der vorliegenden Arbeit wurde eine äußerst einfache Methode entwickelt, die exakte Lösungen von reibungsfreien, axialsymmetrischen Normalkontaktproblemen zwischen starren Indentern und elastisch inhomogenen Medien – so genannten graded materials - erlaubt. Die elastische Inhomogenität ist dabei durch einen Elastizitätsmodul gegeben, der mit der Tiefe des Halbraums nach einem Potenzgesetz zunimmt. Unter der Voraussetzung einer einfach zusammenhängenden Kontaktfläche sind beliebige axialsymmetrische Indenterprofile zulässig. Die Methode kann gleichermaßen zur Lösung von Kontakten mit und ohne Adhäsion angewendet werden. Der kritische Kontaktradius und damit beispielsweise auch die maximale Abzugskraft können aus einer einfachen Forderung an die Geometrie des Ersatzsystems gewonnen werden. Die einfache Anwendung dieser Erweiterung der MDR wurde anhand von Beispielaufgaben demonstriert. Das Spektrum der Aufgaben umfasste einerseits Problemstellungen, deren Lösungen aus sehr aktuellen Untersuchungen hervorgehen und mittels MDR exakt wiedergewonnen wurden; zum Teil wurden diese Lösungen sogar um ergänzende Beiträge bereichert. Letzteres stand aber keinesfalls im Vordergrund. Vielmehr sollte dem Leser die einfache Handhabung der Regeln zur Lösung von Kontaktproblemen nahe gelegt werden. Aus diesem Grunde wurden auch Kontaktprobleme behandelt, deren Lösungen in der Literatur noch nicht zu finden sind. Alle Lösungen wurden dabei auf analytischem Wege berechnet. Es bedarf keines Kommentares, dass komplizierte Kontaktgeometrien natürlich keine analytische Lösung mehr zulassen und numerische Berechnungen notwendig machen. Im Vergleich zur FEM oder BEM, die eine Diskretisierung des Raumes (3D) oder der Oberfläche (2D) verlangen und die Wechselwirkungen zwischen den Elementen berücksichtigen, ist die MDR extrem einfach numerisch implementierbar. Einerseits sind die Elemente nur entlang einer Linie angeordnet (1D) und zum anderen sind die Elemente sogar unabhängig voneinander. Die damit einhergehende enorme Einsparung an Rechenzeit erlaubt den Zugang zu komplexeren Kontaktproblemen.

FEM-Rechnungen von Lee et al. [35] lassen vermuten, dass die gesamten Berechnungen für den power-law graded half-space auch für negative Exponenten der Inhomogenität $-1 \leq k < 0$ gelten. Trotz des unphysikalischen Verhaltens für $z = 0$ und $z \to \infty$ ergaben sich nur geringe Abweichungen im prinzipiellen Lösungsverhalten im Vergleich zu einem mehr realistischem abschnittsweise definierten Gesetz. Es müsste demnach geprüft werden, ob die von Booker et al. [52] gegebene Fundamentallösung für die Oberflächenverschiebung des Halbraums ihre Gültigkeit für $-1 \leq k < 0$ beibehält.

Außerdem stellt sich die Frage, ob sich andere Gesetze auf eine ähnliche Weise mittels MDR umsetzen lassen. Bereits die Überlagerung des Potenzgesetzes mit einem konstanten Anteil wäre diesbezüglich von Interesse. Da beide Anteile getrennt mittels MDR auf Lösungen führen, gilt es zu untersuchen, ob näherungsweise eine gewichtete Superposition der Lösungen realisierbar ist. In diesem Zusammenhang sei der Ansatz von Klein und Duraev [53] genannt, der eine gewichtete Überlagerung der Fundamentallösungen vorsieht.

**Acknowledgements**

The author acknowledges many valuable discussions with Prof. V. L. Popov.

**Appendix A. Ergänzungen zum Kontakt eines konischen Indenters mit parabolischer Spitze**

Die explizite Berechnung der Normalkraft für den Kontakt des inhomogenen Halbraums mit einem konischen Indenter mit parabolischer Spitze sind wir in Abschnitt 3.3 schuldig geblieben. Unter Berücksichtigung von (54) und (55) ergibt sich aus Gleichung (56) für den Bereich $a > b$

$$F_N(a) = 2\frac{h(k,\nu)}{(1-\nu^2)}\frac{E_0}{c_0^k}\left\{\frac{a^{k+1}\delta}{k+1} + \frac{\tan\theta}{b(1+k)(3+k)}\left[(a^2-b^2)^{\frac{k+3}{2}} - a^{3+k}\right] - \frac{\tan\theta}{2(2+k)}B\left(\frac{1}{2},\frac{1+k}{2}\right)\left(a^{2+k} - b^{2+k}\right)\right.$$
$$\left. + \frac{\tan\theta}{2}\frac{\Gamma\left(\frac{-1-k}{2}\right)b^{2+k}}{\Gamma\left(\frac{1-k}{2}\right)}\left[{}_2F_1\left(\frac{1}{2},\frac{-1-k}{2};\frac{3}{2};1\right) - \left(\frac{a}{b}\right)^{1+k}{}_2F_1\left(\frac{1}{2},\frac{-1-k}{2};\frac{3}{2};\frac{b^2}{a^2}\right)\right]\right\}.$$
(100)

Im Grenzfall des homogenen Halbraums $k = 0$ fällt die Lösung zusammen:

$$F_N(a) = \frac{E^* \tan\theta \, a^3}{b}\left[\frac{4}{3} + \frac{b}{a}\arccos\left(\frac{b}{a}\right) - \frac{1}{3}\left(4 - \left(\frac{b}{a}\right)^2\right)\sqrt{1 - \left(\frac{b}{a}\right)^2}\right]. \qquad (101)$$

Für die Eindrücktiefe folgt entsprechend aus Gleichung (55)

$$\delta(a) = \frac{\tan\theta}{b}a^2\left[1 - \sqrt{1 - \left(\frac{b}{a}\right)^2} + \frac{b}{a}\arccos\left(\frac{b}{a}\right)\right]. \qquad (102)$$

Die Normalkraft nach Gleichung (101) und die Eindrücktiefe nach Gleichung (102) für den Eindruck in den homogenen elastischen Halbraum stimmen mit den Lösungen von Ciavarella [47] überein.

**Literaturverzeichnis**

[1]. **Miyamoto, Y., et al.** *Functionally graded materials: design, processing and applications.* Boston / Dordrecht / London : Kluwer Academic Publishers, 1999.

[2]. **Suresh, S.** Graded materials for resistance to contact deformation and damage. *Science's Compass Review.* 2001, Vol. 292, 5526, pp. 2447-2451.

[3]. **Jha, D. K., Kant, T. and Singh, R. K.** A critical review of recent research on functionally graded plates. *Composite Structures.* 2013, 96, pp. 833-849.

[4]. **Fröhlich, O. K.** In Druckverteilung im Baugrunde. Vienna : Springer, 1934, XI Das elastische Verhalten der Böden, pp. 86-108.

[5]. **Borowicka, H.** Die Druckausbreitung im Halbraum bei linear zunehmendem Elastizitätsmodu. *Archive of Applied Mechanics.* 1943, Vol. 14, 2, pp. 75-82.

[6]. **Selvadurai, A. P. S.** The analytical method in geomechanics. *Applied Mechanics Reviews.* 2007, Vol. 60, 3, pp. 87-106.

[7]. **Aleynikov, S.** *Spatial Contact Problems in Geotechnics: Boundary-element Method.* s.l. : Springer Science & Business Media, 2010. pp. 55-83.

[8]. **Booker, J. R., Balaam, N. P. and Davis, E. H.** The behaviour of an elastic non-homogeneous half-space. Part II–circular and strip footings. *International journal for numerical and analytical methods in geomechanics.* 1985, Vol. 9, 4, pp. 369-381.

[9]. **Giannakopoulos, A. E. and Suresh, S.** Indentation of solids with gradients in elastic properties: Part II. Axisymmetric indentors. *International Journal of Solids and Structures.* 1997, Vol. 34, 19, pp. 2393-2428.

[10]. **Chen, S., et al.** Mechanics of adhesive contact on a power-law graded elastic half-space. *Journal of the Mechanics and Physics of Solids.* 2009, Vol. 57, 9, pp. 1437-1448.

[11]. **Jin, F., Guo, X. and Zhang, W.** A unified treatment of axisymmetric adhesive contact on a power-law graded elastic half-space. *Journal of Applied Mechanics.* 2013, Vol. 80, 6, p. 061024.

[12]. **Jin, F., et al.** Adhesive contact of a power-law graded elastic half-space with a randomly rough rigid surface. *International Journal of Solids and Structures.* 2015. doi: 10.1016/j.ijsolstr.2015.12.001.

[13]. **Popov, V. L.** *Kontaktmechanik und Reibung: Von der Nanotribologie bis zur Erdbebendynamik.* Berlin : Springer-Verlag Berlin Heidelberg, 2015. 978-3-662-45974-4.

[14]. **Hertz, H.** Über die Berührung fester elastischer Körper. *Journal für die reine und angewandte Mathematik.* 1882, Vol. 92, pp. 156-171.

[15]. **Cattaneo, C.** Sul contatto di due corpi elastici: distribuzione locale degli sforzi. *Rendiconti dell'Accademia nazionale dei Lincei.* 1938, Vol. 27, pp. 342-348, 434-436, 474-478.

[16]. **Mindlin, R. D.** Compliance of elastic bodies in contact. *Journal of Applied Mechanics.* 1949, Vol. 16, 3, pp. 259–268.

[17]. **Ciavarella, M.** Tangential Loading of General Three-Dimensional Contacts. *Journal of Applied Mechanics.* 1998, Vol. 65, pp. 998-1003.

[18]. **Jäger, J.** Axi-symmetric bodies of equal material in contact under torsion or shift. *Archive of Applied Mechanics.* 1995, Vol. 65, pp. 478-487.

[19]. **Johnson, K. L., Kendall, K. and Roberts, A. D.** Surface energy and the contact of elastic solids. *Proceedings of the Royal Society of London. Series A, Mathematical and Physical Sciences.* 1971, Vol. 324, 1558, pp. 301-313.

[20]. **Radok, J. R. M.** Viscoelastic stress analysis. *Quarterly of Applied Mathematics.* 1957, Vol. 15, pp. 198-202.

[21]. **Popov, V. L.** Deutsch-Russischer Workshop „Numerical simulation methods in tribology: possibilities and limitations. Berlin : s.n., 2005.

[22]. **Geike, T. and Popov, V. L.** Mapping of three-dimenional contact problems into one dimension. *Physical Review E.* 2007, Vol. 76, 3.

[23]. **Heß, M.** *Über die exakte Abbildung ausgewählter dreidimensionaler Kontakte auf Systeme mit niedrigerer räumlicher Dimension.* Berlin : Cuvillier, 2011.

[24]. **Pohrt, R., Popov, V. L. and Filippov, A. E.** Normal contact stiffness of elastic solids with fractal rough surfaces for one-and three-dimensional systems. *Physical Review E.* 2012, Vol. 86, 2. 026710.

[25]. **Popov, V. L. and Heß, M.** *Method of dimensionality reduction in contact mechanics and friction.* Berlin : Springer, 2015.

[26]. **Popov, V. L.** Analytic solution for the limiting shape of profiles due to fretting wear. *Scientific Reports.* 2014, 4. DOI: 10.1038/srep03749.

[27]. **Dimaki, A. V., et al.** Fast High-Resolution Simulation of the Gross Slip Wear of Axially Symmetric Contacts. *Tribology Transactions.* 2015. (just-accepted), 00-00.

[28]. **Willert, E., Heß, M. and Popov, V. L.** Application of the method of dimensionality reduction to contacts under normal and torsional loading. *Facta Universitatis, Series: Mechanical Engineering.* 2015, Vol. 13, 2, pp. 81-90.

[29]. **Persson, B. N. J.** Contact Mechanics for Randomly Rough Surfaces: On the Validity of the Method of Reduction of Dimensionality. *Tribology Letters.* 2015, Vol. 58, 1, pp. 1-4.

[30]. **Popov, V. L.** Comment on "Contact Mechanics for Randomly Rough Surfaces: On the Validity of the Method of Reduction of Dimensionality" by Bo Persson in Tribology Letters. *Tribology Letters.* 2015, Vol. 60, 2, pp. 1-7.

[31]. **Popov, V. L., Pohrt, R. and Heß, M.** General procedure for solution of contact problems under dynamic normal and tangential loading based on the known solution of normal contact problem. 2015. arXiv preprint arXiv:1508.04242.

[32]. **Argatov, I.** A discussion of the method of dimensionality reduction. *Proceedings of the Institution of Mechanical Engineers, Part C: Journal of Mechanical Engineering Science.* 2015, p. 0954406215602512.

[33]. **Popov, V. L. and Heß, M.** Method of dimensionality reduction in contact mechanics and friction: a users handbook. I. Axially-symmetric contacts. *Facta Universitatis, Series: Mechanical Engineering.* 2014, Vol. 12, 1, pp. 1-14.

[34]. **Popov, V. L.** Method of dimensionality reduction in contact mechanics and tribology. Heterogeneous media. *Physical Mesomechanics.* 2014, Vol. 17, 1, pp. 50-57.

[35]. **Lee, D., Barber, J. R. and Thouless, M. D.** Indentation of an elastic half space with material properties varying with depth. *International journal of engineering science.* 2009, Vol. 47, 11, pp. 1274-1283.

[36]. **Mossakovskii, V. I.** Compression of elastic bodies under conditions of adhesion (axisymmetric case). *Journal of Applied Mathematics and Mechanics.* 1963, Vol. 27, 3, pp. 630-643.

[37]. **Jin, Fan and Guo, Xu.** Mechanics of axisymmetric adhesive contact of rough surfaces involving power-law graded materials. *International Journal of Solids and Structures.* 2013, Vol. 50, 20, pp. 3375-3386.

[38]. **Pharr, G. M., Oliver, W. C. and Brotzen, F. R.** On the generality of the relationship among contact stiffness, contact area, and elastic modulus during indentation. *Journal of Materials Research.* 1992, Vol. 7, 3, pp. 613-617.

[39]. **Sneddon, I. N.** The relation between load and penetration in the axisymmetric Boussinesq problem for a punch of arbitrary profile. *International Journal of Engineering Science.* 1965, Vol. 3, pp. 47-57.

[40]. **Borodich, F. M. and Keer, L. M.** Evaluation of elastic modulus of materials by adhesive (no–slip) nano–indentation. *Proceedings of the Royal Society of London A: Mathematical, Physical and Engineering Sciences.* February 2004, Vol. 460, 2042, pp. 507-514.

[41]. **Popov, V. L.** *Contact Mechanics and Friction. Physical Principles and Applications.* Berlin : Springer-Verlag, 2010. pp. 69-70.

[42]. **Awojobi, A. O. and Gibson, R. E.** Plane strain and axially symmetric problems of a linearly nonhomogeneous elastic half-space. *The Quarterly Journal of Mechanics and Applied Mathematics.* 1973, Vol. 26, 3, pp. 285-302.

[43]. **Brown, P. T. and Gibson, R. E.** Surface settlement of a deep elastic stratum whose modulus increases linearly with depth. *Canadian Geotechnical Journal.* 1972, Vol. 9, 4, pp. 467-476.

[44]. **Gibson, R. E.** Some results concerning displacements and stresses in a non-homogeneous elastic half-space. *Geotechnique.* 1967, Vol. 17, 1, pp. 58-67.

[45]. **Rostovtsev, N. A.** An integral equation encountered in the problem of a rigid foundation bearing on nonhomogeneous soil . *Journal of Applied Mathematics and Mechanics.* 1961, Vol. 25, 1, pp. 238-246.

[46]. **Johnson, K. L.** *Contact Mechanics.* s.l. : Cambridge University Press, 1985.

[47]. **Ciavarella, M.** Indentation by nominally flat or conical indenters with rounded corners. *International Journal of Solids and Structures.* 1999, Vol. 36, pp. 4149-4181.

[48]. **Griffith, A. A.** The phenomena of rupture and flow in solids. *Philosophical Transactions of the Royal Society of London, Series A.* 1921, Vol. 221, pp. 163-198.

[49]. —. Theory of rupture. [ed.] Biezeno and Burgers. *Proceedings of the First International Congress for Applied Mechanics.* 1925, pp. 53-64.

[50]. **Heß, M.** On the reduction method of dimensionality: the exact mapping of axisymmetric contact problems with and without adhesion. *Physical Mesomechanics.* 2012, Vol. 15, 4, pp. 19-24.

[51]. **H. Yao, H. Gao.** Optimal shapes for adhesive binding between two elastic bodies. *Journal of colloid and interface science.* 2006, Vol. 298, 2, pp. 564-572.

[52]. **Booker, J. R., Balaam, N. P. and Davis, E. H.** The behaviour of an elastic non-homogeneous half-space. Part I–line and point loads. *International Journal for Numerical and Analytical Methods in Geomechanics.* 1985, Vol. 9, 4, pp. 353-367.

[53]. **Klein, G. K. and Duraev, A. E.** Consinderation of the increase of the modulus of elasticity of soil with increase of depth when calculating beams on a homogeneous supporting soil. *Gidroteh Stroit.* 1971, 7, pp. 19-21.